\documentclass[journal]{vgtc}              

\DeclareGraphicsExtensions{.png,PNG,.pdf,.PDF}

\onlineid{1645}

\vgtccategory{Research}

\def \sys {{HybridLens}}

\def \etal {{\emph{et al}.\thinspace}}

\def \eg {{\emph{e.g}.}}
\def \ie {{\emph{i.e}.}}

\title{Interactive Hybrid Rice Breeding with Parametric Dual Projection}

\author{%
  Changjian Chen,
  Pengcheng Wang,
  Fei Lyu,
  Zhuo Tang,
  Li Yang,
  Long Wang,
  Yong Cai,
  Feng Yu,
  and Kenli Li
}

\authorfooter{
  \item
    All the authors are from Hunan University.
    \\E-mail: \{changjianchen, wangpengcheng, feilv, ztang, yanglixt, wanglong8591, caiyong911, feng\_yu, lkl\}@hnu.edu.cn. C.~Chen and P.~Wang are joint first authors. Z.~Tang is the corresponding author.
}

\abstract{
Hybrid rice breeding crossbreeds different rice lines and cultivates the resulting hybrids in fields to select those with desirable agronomic traits, such as higher yields.
Recently, genomic selection has emerged as an efficient way for hybrid rice breeding. 
It predicts the traits of hybrids based on their genes, which helps exclude many undesired hybrids, largely reducing the workload of field cultivation.
However, due to the limited accuracy of genomic prediction models, breeders still need to combine their experience with the models to identify regulatory genes that control traits and select hybrids, which remains a time-consuming process.
To ease this process, in this paper, we proposed a visual analysis method to facilitate interactive hybrid rice breeding.
Regulatory gene identification and hybrid selection naturally ensemble a dual-analysis task.
Therefore, we developed a parametric dual projection method with theoretical guarantees to facilitate interactive dual analysis.
Based on this dual projection method, we further developed a gene visualization and a hybrid visualization to verify the identified regulatory genes and hybrids. 
The effectiveness of our method is demonstrated through the quantitative evaluation of the parametric dual projection method, identified regulatory genes and desired hybrids in the case study, and positive feedback from breeders.

}


\teaser{
  \centering
  \includegraphics[width=1.0\linewidth]{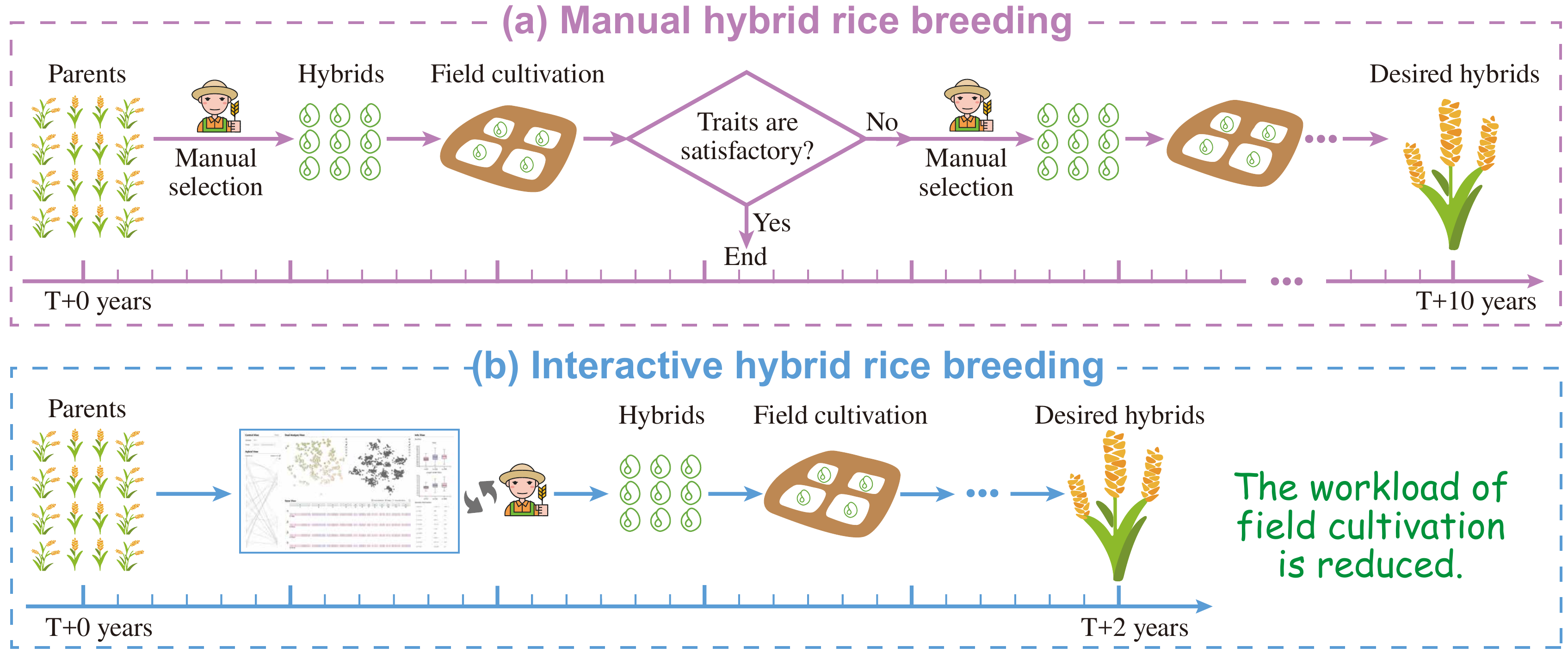}
  \caption{The comparison between the manual and the proposed interactive hybrid rice breeding methods.
  (a) The manual hybrid rice breeding relies heavily on the experience of breeders, involving numerous repetitions and usually spanning over ten years.
  (b) The interactive hybrid rice breeding combines the experience of breeders with genomic data and prediction models, which reduces the workload of field cultivation.
  }
  \label{fig:teaser}
}

\graphicspath{{figs/}{figures/}{pictures/}{images/}{./}} 

\usepackage{multirow}
\usepackage{color}
\usepackage{lipsum}                    
\usepackage{amssymb,amsmath}
\usepackage{bbm} 
\usepackage{tabularx} 
\usepackage{wrapfig}
\usepackage{booktabs}
\usepackage{makecell}
\usepackage{graphicx}
\usepackage{dblfloatfix}

\newcommand{\myparagraph}[1]{\vspace{1mm}\noindent\textbf{#1}}

\newcommand{\pcheng}[1]{\textcolor{black}{#1}}
\newcommand{\changjian}[1]{\textcolor{black}{#1}}

\usepackage{mathptmx}                  

\begin{document}


\firstsection{Introduction}

\maketitle
\fontsize{9}{9} 

\begin{figure}[t]
    \centering
    \includegraphics[width=\linewidth]{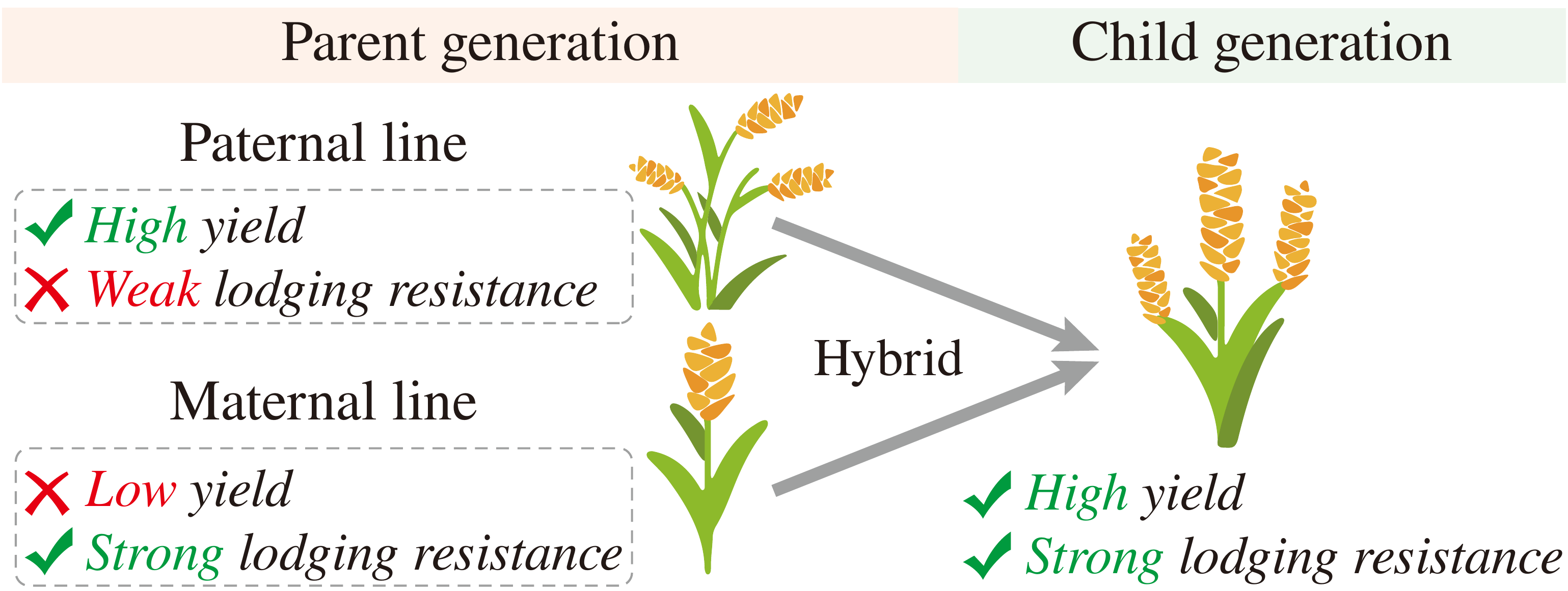}
    \caption{Illustration of hybrid rice breeding.}
    \label{fig:hybrid-example}
\end{figure}

The improvement of rice yields and qualities through hybrid breeding has long been a cornerstone of global food security~\cite{yuan2017progress}.
A hybrid is the child of two rice lines, referred to as the paternal and maternal lines (\eg, Fig.~\ref{fig:hybrid-example}).
By crossbreeding different rice lines and careful selection, the rice hybrids can gain better agronomic traits, such as yields and lodging resistance.
Conventionally, to obtain better hybrids, breeders manually select different rice lines for crossbreeding according to their experience, cultivate them in fields, and collect their agronomic traits.
If the traits of the hybrids are unsatisfactory, they explore other rice lines and repeat the process, which usually takes more than ten years (\eg, Fig.~\ref{fig:teaser}(a))~\cite{xu2021genomic}.
Recently, due to the advancements in molecular techniques, genomic selection has emerged as an efficient way for hybrid breeding.
In relatively stable environments, the traits of rice hybrids are mainly determined by their genes.
Therefore, in genomic selection, a genomic prediction model is utilized to predict the traits of hybrids based on their genes.
Then, the hybrids with desired traits can be selected without expensive and labor-intensive field cultivation.

In practice, due to limited data and annotation errors, the accuracy of genomic prediction models is usually constrained~\cite{wang2023dnngp}, which makes the genomic selection less effective.
A more practical way is to integrate genomic selection with field cultivation.
Breeders first utilize genomic prediction models to predict the traits of hybrids.
Then, they leverage their experience, along with prediction results and genomic data, to select hybrids for field cultivation.
Although this method reduces the workload of field cultivation, two challenges still exist.
1) \textbf{Regulatory gene identification}. 
Regulatory genes regulate traits, which helps select paternal and maternal lines for crossbreeding.
Although many studies have explored regulatory genes that regulate rice traits, the effects of regulatory genes vary across different rice populations~\cite{wei2021quantitative}.  
Therefore, it is necessary to analyze the effects of genes based on the data from the target rice population.
However, due to the large volume of genes, quickly identifying the regulatory ones remains a challenge.
2) \textbf{Hybrid selection}. 
Selecting hybrids requires the simultaneous integration and analysis of prediction results, genomic data, and breeder experience.
This presents challenges for breeders, as it is difficult to analyze all of them concurrently~\cite{wang2016topicpanorama}.
Therefore, a tool that facilitates efficient exploration and analysis is required.


To address these challenges, we develop {\sys} (Fig.~\ref{fig:teaser}(b)), a visual analysis method to help 1) explore the relationships between hybrid and genomic data efficiently and 2) facilitate the selection of desired hybrids for field cultivation. 
Our method is based on a key observation: the regulatory gene identification involves identifying key genes based on the analysis of the hybrids, 
and the hybrid selection involves identifying desired hybrids based on the analysis of the genes. 
This makes them naturally a \textbf{dual-analysis task}.
Therefore, we propose using the dual projection method to facilitate both the regulatory gene identification and the hybrid selection.
Although there are existing methods, such as WMDS~\cite{endert2011observation} and SIRIUS~\cite{dowling2018sirius}, for dual projection, we found that they cannot meet the interactive requirement, and the neighborhood relationships are not well-preserved.
To this end, we developed a parametric dual projection method based on invert networks~\cite{jing2021hinet}.
We theoretically and experimentally prove that our parametric dual projection method is faster and better \changjian{in preserving neighborhood relationships} than the existing methods.
Based on this dual projection method, we developed a gene visualization to identify the regulatory genes that regulate the traits of interest, and a hybrid visualization to select
hybrids with desired traits.
The effectiveness of our method is demonstrated through a quantitative evaluation of the dual projection method, identified regulatory genes and selected hybrids in the case study, and positive feedback from breeders. \pcheng{The source code is available at: https://github.com/hnu-vis/ParametricDualProjection.}

In summary, the contributions of this work include:

\begin{itemize}[nosep]
\item\noindent{\textbf{A parametric dual projection method} \changjian{for dual analysis, which is domain-agnostic and} theoretically and experimentally better than the existing methods. 
} 
\item\noindent{\textbf{A visual analysis tool} that supports the identification of regulatory genes and the selection of hybrids for better traits.}
\item\noindent{\textbf{A case study} with the real-world data from Hunan Province, China, to demonstrate the effectiveness of the proposed method.}

\end{itemize}

\section{Related Work}

Our work focuses on the interactive analysis of the relationships between hybrid and genomic data.
The most relevant topics in the visualization field are genomic visualization and interactive dual analysis.

\subsection{Genomic visualization}
Genomic visualization methods aim to facilitate the exploration and analysis of genomic data through visualizations.
Based on how the visualizations are developed, these methods can be classified into two categories: manual construction and semi-autonomic construction.

\textbf{Manual construction} focuses on manually designing and developing tailored genomic visualizations to meet the requirements of biological experts.
Thorvaldsd{\'o}ttir~\etal~\cite{thorvaldsdottir2013igv} proposed the integrated genomic viewer, which supports the exploration of large genomic datasets at different levels of detail.
Thereafter, many works introduced visual components and interactions to enhance the exploration.
Nguyen~\etal~\cite{nguyen2016visual} introduced similarity space construction and gene-to-gene comparison to help analyze cancer genes in different patients. 
\pcheng{Nusrat~\etal~\cite{nusrat2019tasks} surveyed genomic data visualization, proposing taxonomies for data, techniques, and tasks for genomic data analysis.}
PanVA~\cite{astrid2023panva} provides various interaction methods, such as grouping and aggregation, allowing users to explore data relations from different perspectives.

\textbf{Semi-autonomic construction} focuses on improving the efficiency of constructing genomic visualizations. 
A pioneering work along this line is MGV~\cite{kerkhoven2004mgv}, which allows interactively select genomic visualizations, such as chromosome wheels and linear genome maps. 
SeqCode~\cite{blanco2021seqcode} further extended this idea to support large-scale genomic data.
When facing complex tasks, selecting appropriate visualizations still costs a lot of human effort.
To tackle this challenge, GenoREC~\cite{pandey2022genorec} recommends suitable genomic visualizations based on descriptions of the data and analysis tasks.
All the methods mentioned above are limited to predefined visualizations.
To help customize visualizations efficiently, many recent studies~\cite{lyi2021gosling, l2022multi, wang2023enabling, tang2024jcvi} have focused on developing programming tools.
LYi~\etal~\cite{lyi2021gosling} developed Gosling, a grammar for interactive and scalable genomic visualization, and a library based on it for efficient visualization rendering.
Tang~\etal~\cite{tang2024jcvi} developed JCVI, a Python‐based library to support both genomic visualization creation and broad genomic tasks, such as genome assemblies and annotations.

Though effective, these methods mainly focus on exploring genomic data. 
However,  to support hybrid rice breeding, it is necessary to explore both hybrid and genomic data and their relationships, which are not supported by the existing methods.
To address this gap, we developed a dual projection method that enables a more effective exploration of the relationships between hybrid and genomic data.


\subsection{Interactive Dual Analysis}

Interactive dual analysis enhances the exploration of high-dimensional data through joint analysis of attributes~\cite{dennig2023fs}.
The existing methods can be classified into selection-based and weighting-based methods.

\textbf{Selection-based} dual analysis methods allow users to select a subset of data/attributes for further analysis.
Brushing and linking is the most simple strategy for selection-based dual analysis.
The existing works utilize this strategy to reveal the data-attribute interactions (~\cite{artur2019novel},~\cite{fernstad2013quality},~\cite{muller2021integrated},~\cite{turkay2014attribute},~\cite{van2016exploring}), refine the downstream model performance by iteratively selecting attributes (~\cite{rauber2018projections},~\cite{zhao2019featureexplorer}), achieve progressive analysis by integrating an online algorithm~\cite{turkay2016designing}, and analyze cancer subtypes by exploring linked views~\cite{turkay2014characterizing}.
Some recent studies extend the brushing and linking strategy to support large-scale datasets.
Turkay~\etal~\cite{turkay2012representative} summarizes representative factors for the selected attributes to enable in-depth analysis. 
Yuan~\etal~\cite{yuan2013dimension} divides the data into subsets and proposes a dimension project tree for multi-level data exploration. 
DimLift~\cite{garrison2021dimlift} groups the attributes with similar data contributions together, facilitating hierarchical attribute exploration.
 
Although selection-based dual analysis methods are widely used, they require multiple selection operations, placing high demands on users.
Therefore, weighting-based dual analysis methods are proposed.

\textbf{Weighting-based} dual analysis methods enable users to modify what they see in the visualizations. 
Then, the underlying weights of the data/attributes are updated automatically to accommodate the modifications.
This concept is first introduced by Endert~\etal~\cite{endert2011observation}.
They developed WMDS to automatically update the weights of the attributes once users modify the positions of the data.
Self~\etal~\cite{self2016bridging} extended WMDS to non-expert-friendly interactions, including moving, expanding/contracting, and adjusting weights for data points, simplifying interaction and improving the user experience. 
Dowling~\etal~\cite{dowling2018sirius} proposed SIRIUS, a symmetric dual projection technique that enables modification to both projections through interactions in one, helping users to analyze both data and attributes effectively.

The weighting-based methods help users explore and analyze the relationships between data and attributes effectively.
However, they are computationally expensive and cannot meet the interactive requirements when the dataset is large.
To tackle this issue, we developed a parametric dual projection method, which is fast enough to meet the interactive requirements and preserve neighborhood relationships more effectively than existing weighting-based methods.

\begin{figure*}[b]
    \centering
    \includegraphics[width=\linewidth]{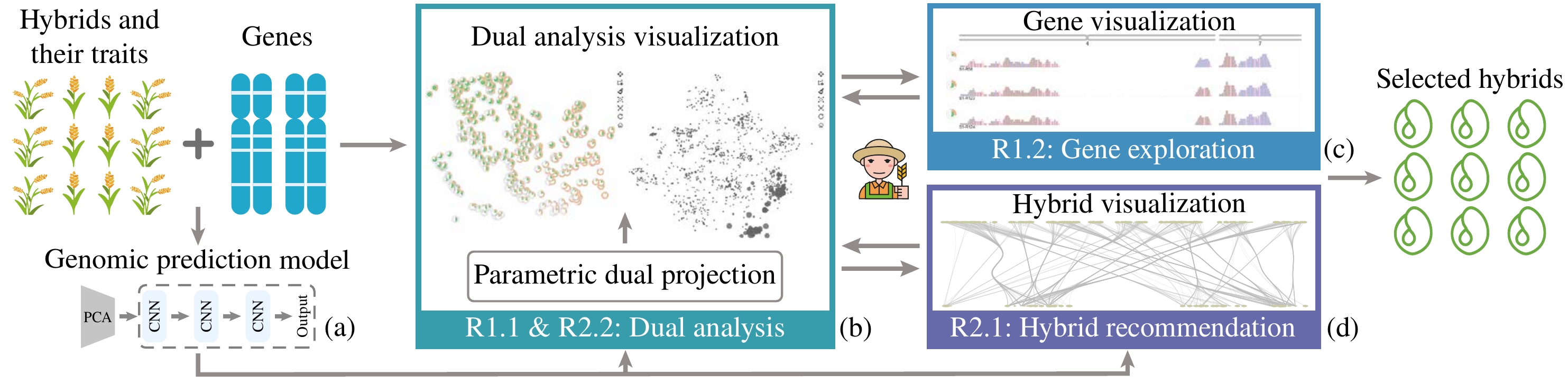}
    \caption{{\sys} overview: given a set of rice hybrids, their genes, and the traits, (a) a genomic prediction model is trained; (b)-(d) three coordinated visualizations are provided to help explore the relationships between hybrids and genes and select hybrids with desired traits. After the hybrids are selected, breeders cultivate them in the fields to verify their traits.}
    \label{fig:pipeline}
\end{figure*}

\begin{figure}
    \centering
    \includegraphics[width=\linewidth]{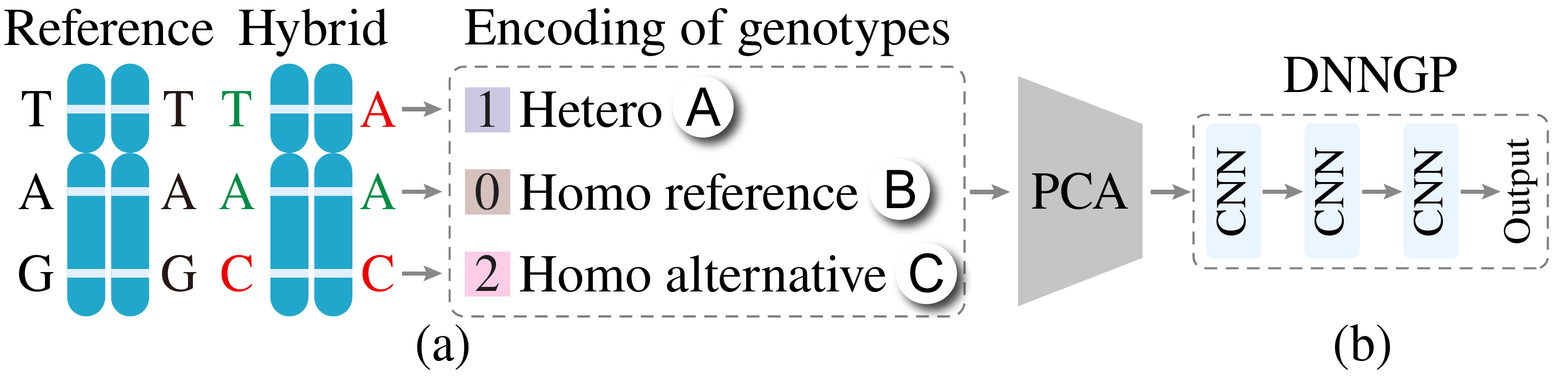}
    \caption{Genomic selection consists of two main steps: (a) data processing and (b) genomic prediction model training.}
    \label{fig:background}
    \vspace{-1mm}
\end{figure}

\section{Background: Genomic Selection}
\label{sec:background}
Genomic selection utilizes genomic data to predict the traits of hybrids and select the desired ones.
It involves two key steps: data processing and genomic prediction model training.

\myparagraph{Data processing}.
The data processing step converts the genomic data to numerical vectors that can be processed by genomic prediction models.
Following the common practice~\cite{gu2023structure}, this process uses a well-recognized rice line as the reference, such as Nipponbare~\cite{kawahara2013improvement}.
For each genomic locus on the chromosomes of a hybrid, the encoding of genotypes is determined as follows: if the locus is heterozygous, it is encoded as one (hetero gene, Fig.~\ref{fig:background}A);
if the locus is homozygous and identical to that of the reference, it is encoded as zero (homo reference gene, Fig.~\ref{fig:background}B);
if the locus is homozygous but differs from that of the reference, it is encoded as two (homo alternative gene, Fig.~\ref{fig:background}C).
With such a strategy, the genomic data of a hybrid is converted into a numerical vector.
It is important to note that the term for the genomic locus is Single Nucleotide Polymorphism. 
However, due to the poor readability of this term, we use the term ``gene'' to refer to Single Nucleotide Polymorphism in this paper.

\myparagraph{Genomic prediction model training}.
After data processing, we utilize DNNGP~\cite{wang2023dnngp}, the state-of-the-art genomic prediction model, for training. 
DNNGP is a neural network with three one-dimensional CNN layers, which can predict multiple traits simultaneously. 
Since the dimensionality of the processed genomic data is usually very large, similar to DNNGP, we first apply PCA to reduce their dimensionality to 200. 
Then, the dimensionally reduced data is used for training.

\section{Requirement Analysis}
\label{sec:requirement}

This work was developed in close collaboration with four rice breeders (B1-B4).
B1 and B2 are breeders from the research academy of  Longping High-Tech Agriculture Co., Ltd., which is one of the biggest rice seed companies in the world.
Both B1 and B2 possess over ten years of experience in manual hybrid rice breeding. 
Currently, they have struggled with inefficiencies in manual hybrid selection and are eager to utilize genomic prediction models to assist them in breeding.
B3 and B4 are researchers from the Hunan Hybrid Rice Research Center, founded by the ``Father of Hybrid Rice,'' Longping Yuan, and globally recognized as a leading institution in hybrid rice research.
They are dedicated to developing advanced genomic prediction models.
However, they have faced challenges in improving the accuracy of these models.
Thus, they are exploring ways to better utilize them to help hybrid selection with limited accuracy.
The collaboration spanned over 14 months.
We conducted interviews with them every 1-2 months to collect requirements and feedback on our developed prototypes. 
Based on this, we have summarized the following key requirements.

\myparagraph{R1: Identifying the regulatory genes that regulate the traits of interest}.
All breeders stated that an early step in hybrid rice breeding is to identify the regulatory genes that regulate the traits of interest.
``If a regulatory gene leads to better traits, I tend to select the hybrids with this gene for field cultivation because they are more likely to have the desired traits.'' B1 said.
Although many studies have identified regulatory genes that regulate traits, their effects can differ in different rice populations.
``For example, the gene bZIP73 makes the Indica rice sensitive to low temperatures, whereas it causes Japonica rice to be tolerant to low temperatures~\cite{liu2019bZIP73}.'' B3 said.
Therefore, breeders still need to explore the data from the target rice population to identify the regulatory genes that regulate the traits of interest.

\textbf{R1.1:} \textit{Exploring how genes regulate the traits of hybrids to identify regulatory ones}.
Currently, the breeders use the correlation between a gene and a type of trait to analyze how the gene regulates the trait.
This method is effective when a type of trait is regulated by a single gene~\cite{huang2010genome}.
However, in practice, a type of trait (\eg, yield) is often regulated by multiple genes,  and a single gene can regulate multiple traits.
The correlation-based methods usually fail under these circumstances.
Therefore, all the breeders expressed a strong interest in analyzing the complex relationships between genes and the traits of hybrids.

\textbf{R1.2:} \textit{Analyzing the regulatory genes in the context of neighbors on chromosomes}.
Both B3 and B4 pointed out that genes regulating the same traits were usually located at neighboring genomic loci on chromosomes.
Examining these genes in the context of neighbors on chromosomes helps analyze multiple genes together, which saves their efforts.
Moreover, B2 highlighted that it was important to compare the genes of different hybrids at the same genomic loci.
This helps him confirm whether the identified regulatory genes regulate the traits of interest.
\looseness=-1

\myparagraph{R2: Selecting hybrids with desired traits for field cultivation}.
The breeders pointed out that selecting hybrids for field cultivation involves considering multiple constraints. 
First, it is necessary to choose paternal and maternal lines that have relatively high genetic distances, which usually leads to better traits according to their experience and existing studies~\cite{gu2023structure, lv2020resequencing}.
Second, the breeders often need to consider multiple traits simultaneously. 
For example, for Indica rice, a higher length/width ratio enhances eating quality, but it is also crucial to maintain a sufficient yield to ensure economic value.
Moreover, since the genomic prediction model is not highly reliable, the breeders still need to combine it with the regulatory genes for hybrid selection.

\textbf{R2.1:} \textit{Identifying hybrid candidates under multiple constraints}.
Currently, the breeders must consider all these constraints simultaneously when selecting hybrids, which is both inefficient and ineffective. 
They would prefer to have some hybrid candidates recommended to start with.
Moreover, they also want to explore how well the recommended hybrid candidates meet these constraints.
This helps them determine whether these candidates are worth further investigation.

\textbf{R2.2:} \textit{Analyzing the hybrid candidates combining both the genomic prediction results and regulatory genes}.
After obtaining the hybrid candidates, directly using genomic prediction results to select those with the desired traits may not be reliable due to the limited accuracy of these models. 
Therefore, the breeders hoped to combine genomic prediction results with regulatory genes to jointly assess the traits of the hybrid candidates, thereby identifying the ones with desired traits.

\section{{\sys} Visualization}
\label{sec:method}
Based on the identified requirements, we developed {\sys} to support interactive hybrid rice breeding.
Fig.~\ref{fig:pipeline} provides an overview of the developed method.
Given a set of rice hybrids, their genes, and the traits collected through field cultivation, a genomic prediction model is trained (Fig.~\ref{fig:pipeline}(a)). 
Then, the hybrids and their traits, genes, and prediction results are fed into the dual analysis visualization to help explore the relationships between hybrids and genes (\textbf{R1.1} and \textbf{R2.2}, Fig.~\ref{fig:pipeline}(b)).
Based on the dual analysis, users utilize the gene visualization to identify the regulatory genes that regulate the traits of interest (\textbf{R1.2}, Fig.~\ref{fig:pipeline}(c)), and the hybrid visualization to select hybrids with desired traits (\textbf{R2.1}, Fig.~\ref{fig:pipeline}(d)).
These selected hybrids will be cultivated in fields by breeders to verify whether their traits are desired.

\subsection{Dual Analysis Visualization}
To facilitate the identification of regulatory genes and hybrid selection, it is essential to explore and analyze both the hybrids and genes, as well as their relationships.
\changjian{According to the survey of Dennig~\etal~\cite{dennig2023fs}, the most common way is to utilize projection methods,}
such as t-SNE~\cite{van2008tsne} and PCA~\cite{wold1987pca}, to project them into the two-dimensional plane and present them as scatterplots.
Although this strategy can support the exploration of hybrid and genes well, it fails to explain their relationships~\cite{dowling2018sirius,yang2024foundation,lyu2024supercomputer}. 

To facilitate the exploration of hybrids and genes and their relationships, inspired by the recent work SIRIUS~\cite{dowling2018sirius}, we proposed utilizing the dual projection method.
This method projects hybrids and genes into two interconnected scatterplots.
When users modify one scatterplot, the other scatterplot will update to reflect the modification, which reveals the relationships between them. 
\pcheng{For example, in the hybrid scatterplot, users observe that hybrids with low and high yields are placed in separate regions. 
To figure out which genes regulate yield and cause this separation, users move hybrids with low yields to regions of high yields.
Upon update, the weights of certain genes in the gene scatterplot increase, indicating the genes that regulate yield.
}
Building on this idea, we developed a dual analysis visualization supported by a parametric dual projection method.

\subsubsection{Dual projection}

\myparagraph{Problem setting.}
The input of the dual projection is the hybrid data $X$ processed by the data processing method described in Sec.~\ref{sec:background}.
Each row $x_i$ represents a hybrid.
Accordingly, $X^{T}$, the transpose of $X$, represents the genes, where each row represents a gene.
The dual projection is to find two projection functions $p_s(\cdot)$ and $p_g (\cdot)$ 
that map $X$ and $X^{T}$ to a two-dimensional plane:
\begin{align}
    S &= p_s(X) \ \ \ \ \mathrm{where} \ \ p_s = \underset{p_s}{\arg\min} \ O_s(p_s, X) \label{eq:hybrid-proj},\\
    G &= p_g(X^T)\ \ \mathrm{where} \ \ p_g = \underset{p_g}{\arg\min} \ O_g(p_g, X^T). \label{eq:gene-proj}
\end{align}
The projection functions can be obtained by any projection method, such as t-SNE.
$O_s(\cdot, \cdot)$ and $O_g(\cdot, \cdot)$ are the objective functions of the used projection method.
$S=\{s_1, s_2, ...\}$ and $G=\{g_1, g_2, ...\}$ are the positions of hybrids and genes in the two-dimensional plane, respectively. 
\looseness=-1

When users modify one scatterplot, the other scatterplot will update accordingly.
The previous studies~\cite{dowling2018sirius, endert2011observation} show that a practical way to achieving this involves a two-step process.
Take modifying $S$ to $S'$ as an example, this process is formulated as:
\begin{equation}
    \begin{aligned}
    \label{eq:x_update}
        X' & = &  \underset{X} {\arg\min} & & &   \|p_s(X) - S'\|^2 \\
           & & \mathrm{s.t.}  & & &  p_s = \underset{p_s}{\arg\min} O_s(p_s, X).
    \end{aligned}
\end{equation}
\begin{equation}
    \begin{aligned}
    \label{eq:g_update}
        G' = p_g(X'^T) \ \ \mathrm{where} \ \ p_g = \underset{p_g}{\arg\min} O_g(p_g, X'^T).
    \end{aligned}
\end{equation}
Here, the first step (Eq.~(\ref{eq:x_update})) determines an updated $X'$ that reflects the modification.
Then, the other scatterplot is updated based on $X'$ (Eq.~(\ref{eq:g_update})).
The scatterplot can be updated in the same way when modifying $G$ to $G'$.
Since the updating strategies for the modifications of $S$ and $G$ are identical, we only use the modification of $S$ as an example to illustrate the basic idea in the following sections.

\begin{figure}[b]
    \centering
    \includegraphics[width=\linewidth]{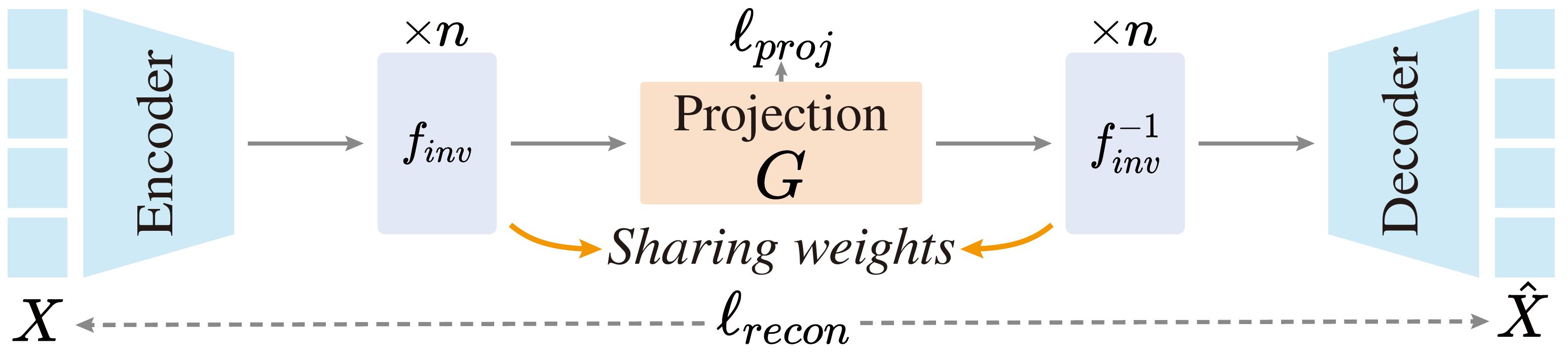}
    \caption{The structure of the parametric dual projection method.}
    \label{fig:model}
\end{figure}


\myparagraph{Existing dual projection method}.
The key to updating the dual projection is optimizing Eq.~(\ref{eq:x_update}).
Eq.~(\ref{eq:x_update}) is a bi-level optimization~\cite{bard2013practical}, which has to solve the projection problem under the constraint many times.
This causes a huge computational cost.
To address this issue, the existing state-of-the-art (SOTA) dual projection method, SIRIUS~\cite{dowling2018sirius}, assumes the updated $X'$ is obtained by reweighting the genes, and the projection method is MDS~\cite{gansner2005graph}.
Then, the Eq.~(\ref{eq:x_update}) can be simplified as $X' = XW$, where
$W$ is obtained by:
\begin{equation}
\begin{aligned}
W = \underset{W}{\arg\min} \sum_{ij}& \left\| \sqrt{(x_i W - x_j W)^2} - \sqrt{(s_i' - s_j')^2} \right\|^2. \label{eq:sirius}
\end{aligned}
\end{equation}
The detailed derivation can be found in the supplemental material.
$W$ is a diagonal matrix where the $i$-th element is the weight of the $i$-th gene.
This problem can be solved by the stress majorization algorithm~\cite{endert2011observation}.

SIRIUS reduces the bi-level optimization into a single-level optimization, largely reducing the computational cost.
However, Eq.~(\ref{eq:sirius}) does not have a closed-form solution and needs to be optimized iteratively, which still cannot meet interactive requirements (less than one second). 
For example, given a dataset with 714 hybrids and 2,081 genes, updating the scatterplots takes approximately 17 seconds.
Additionally, this method is limited to using MDS, which fails to preserve neighborhood relationships effectively, hindering the exploration of hybrids and genes.

\begin{figure*}[b]
    \centering
    \includegraphics[width=\linewidth]{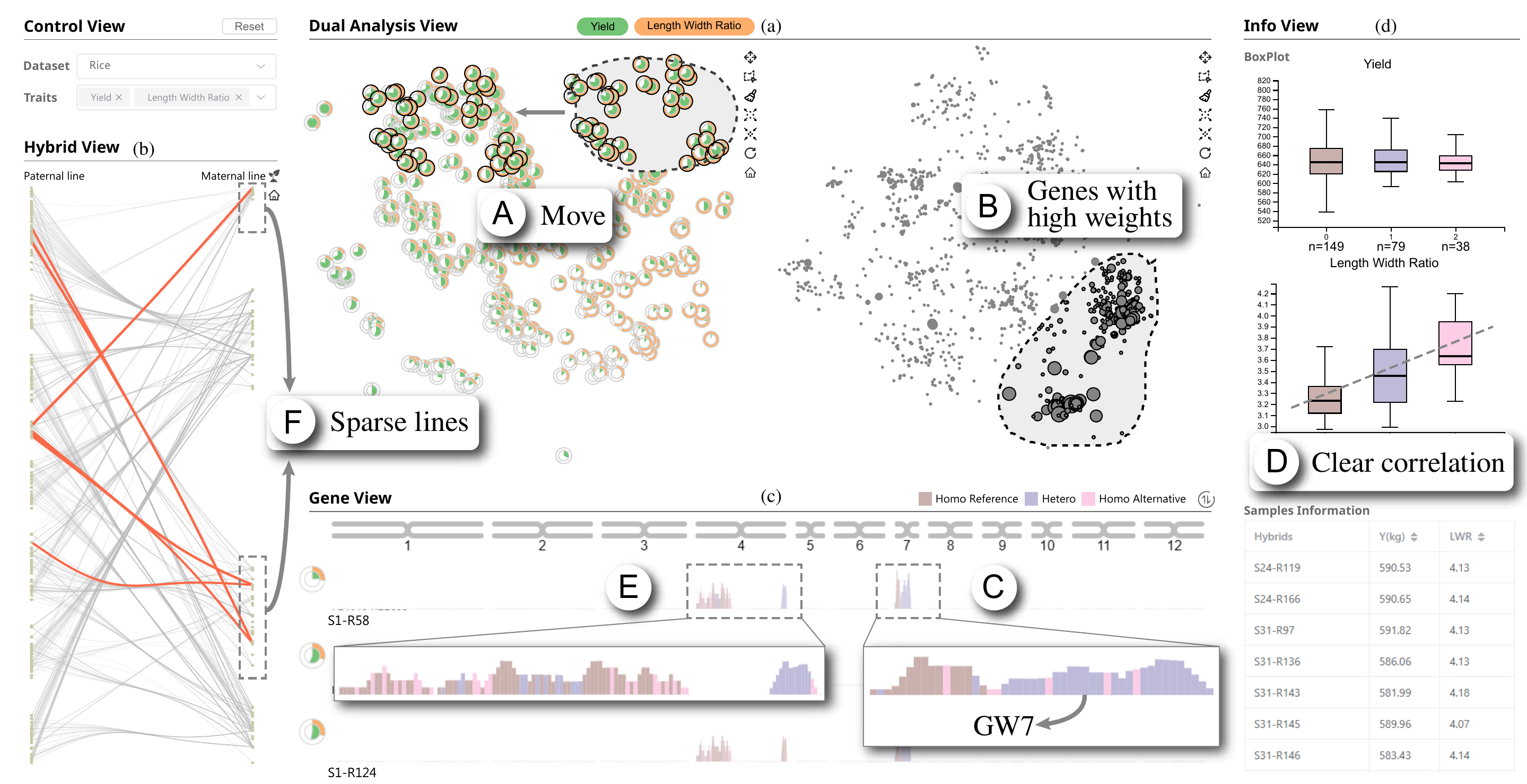}
    \caption{{\sys}: (a) a dual analysis visualization facilitates the exploration of the relationships between genes and hybrids; (b) a hybrid visualization helps select hybrids with desired traits; (c) a gene visualization helps identify the regulatory genes; (d) an information panel to show how the trait values
are distributed among different genotypes of a selected gene.}
    \label{fig:system}
\end{figure*}


\myparagraph{Our parametric dual projection method}.
To tackle these two issues, we get inspiration from the invertible neural networks~\cite{kingma2018glow}.
A main advantage of an invertible neural network $f_{inv}$ is that its input $X$ can be obtained from its output by a single backward pass of the network, namely, $X=f_{inv}^{-1}(S)$.
Here, $S = f_{inv}(X)$.
If we use an invertible neural network as the projection function, we can prove the following theorem. 
\newtheorem{theorem}{Theorem}
\begin{theorem}
\label{theorem:opt}
Using an invertible neural network as the projection function $p_s(\cdot)$. When $S$ is modified to $S'$, let $X'_{\text{opt}}$ be the optimal updated X obtained by optimizing Eq.~(\ref{eq:x_update}).
Then we have
$$X'_{\text{opt}} = p_s^{-1}(S').$$
\end{theorem}
A formal proof of this theorem is given in the supplemental material.
This theorem implies that by using invertible neural networks as the projection functions, Eq.~(\ref{eq:x_update}) can be optimized by a single backward pass of the networks. 
With modern invertible neural networks, such a backward can be finished in one second within a GPU~\cite{kingma2018glow}, which makes it meet the interactive requirements.
Moreover, a recent work, CDR~\cite{xia2022cdr}, shows that neural-network-based projection methods achieve SOTA projection performance.
Therefore, we developed a parametric dual projection method based on the invertible neural networks.

Here, we take the projection of hybrids as an example to explain the main idea. 
The projection of genes can be obtained similarly.
As shown in Fig.~\ref{fig:model}, our method utilizes invertible neural networks nested in an autoencoder for dual projection.
The autoencoder is introduced to enable the inverse process in the latent space.
This is inspired by Stable Diffusion~\cite{rombach2022high}, which enables the diffusion process in the latent space to mitigate overfitting.
With this network, the training process is to preserve neighborhood relationships while ensuring the autoencoder can reconstruct the input:
\begin{equation}
    \mathcal{L} = \ell_{proj} + \lambda\ell_{recon}.
\end{equation}
The first term ensures that neighborhood relationships are preserved.
The second term ensures that the autoencoder can reconstruct the data.
$\lambda$ is the weight to balance the two terms and is determined by the multi-task learning method~\cite{Liu_Liang_Gitter_2019, chen2024enhancing, zhou2025hierarchical}. 

For the first term $\ell_{proj}$, the contrastive loss is utilized, which has been proven to be effective in preserving neighborhood relationships~\cite{xia2022cdr}:
\begin{equation}
    \ell_{proj} = -\log \frac{\ \exp(\mathrm{sim}(x_i, x_j) / \tau)}{\sum_{t=1}^{2B} \mathbbm{1}_{[t \neq i]} \ \exp(\mathrm{sim}(x_i, x_t) / \tau)}.
\end{equation}
$B$ is the size of a mini-batch during training.
$\mathbbm{1}_{[t \neq i]}=1$ if $t \neq i$; otherwise $\mathbbm{1}_{[t \neq i]}=0$. 
$\mathrm{sim}(\cdot, \cdot)$ is the similarity between two hybrids, and $\tau$ is the temperature parameter.

For the second term $\ell_{recon}$, to ensure the autoencoder can reconstruct the input with the help of invertible neural networks, the loss is defined as follows:
\begin{equation}
    \ell_{recon} = \sum_i(x_i-\hat{x}_i)^2.
\end{equation}
Here, $\hat{X}=\{\hat{x}_1, \hat{x}_2, ...\}$ denotes the reconstructed output.

Once the neural network is trained, the projection function is defined as: $S = p_s(X) = f_{inv}(\mathrm{Encoder}(X))$.
When the $S$ is modified to $S'$, $X$ and $G$ are updated accordingly: $X' = p_s^{-1}(S') = \mathrm{Decoder}(f_{inv}^{-1}(S'))$, $G = p_g(X'^T)$.
With this network, we can prove the following theorem.
\begin{theorem}
\label{theorem:comparison}
Using an invertible neural network as the projection method $p_s(\cdot)$. When $S$ is modified to $S'$, let $X'_{\text{SIRIUS}}$ be the updated X obtained by SIRIUS, and $X'_{\text{inv}}$ be the updated X obtained by our method.
Then we have 
$$\|p_s(X'_{inv}) - S'\|^2 \leq \|p_s(X'_{SIRIUS}) - S'\|^2.$$
\end{theorem}
A formal proof of this theorem is given in the supplemental material.
This theorem implies that the proposed parametric projection method can always be better than the existing SOTA method, SIRIUS, \pcheng{in updating the other scatterplot when one scatterplot is modified}.

\begin{table*}[b]
\centering
\caption{Comparison of single and dual projection methods. The best result in each setting is in bold.}
\begin{tabular}{c|c|ccc|ccc|ccc}
\hline
 & \multirow{2}{*}{Methods} & \multicolumn{3}{c|}{MNIST} & \multicolumn{3}{c|}{CIFAR-10}& \multicolumn{3}{c}{Genomic dataset} \\ \cline{3-11} 
 &                          & T(30) & C(30) & Time & T(30) & C(30) & Time & T(30) & C(30) & Time \\ \hline
\multirow{2}{*}{Single} 
 & t-SNE                    & \textbf{0.963} & \textbf{0.950} & / & \textbf{0.951} & \textbf{0.957}& / & \textbf{0.957} & \textbf{0.948} & / \\ 
 & PCA                      & 0.734 & 0.904 & / & 0.784 & 0.917 & / & 0.856 & 0.910 & / \\ 
  \hline
\multirow{2}{*}{Dual}   
 & SIRIUS                   & 0.781 & 0.873 &  581.86& 0.784 & 0.900 &  1430.4& 0.888 & 0.919 &  17.23\\
 & Ours                     & \textbf{0.961} & \textbf{0.944} &  \textbf{0.01}& \textbf{0.935} & \textbf{0.958} &  \textbf{0.01}& \textbf{0.966} & \textbf{0.950} &  \textbf{0.01}\\ \hline
\end{tabular}
\label{table-experiments}
\end{table*}

\subsubsection{Dual analysis}
With the parametric dual projection method, we visually present the hybrids and genes in scatterplots, allowing users to explore the relationships between hybrids and genes interactively.

\myparagraph{Visual encoding}.
The hybrids and genes are presented in scatterplots (Fig.~\ref{fig:system}(a)).
In the \textit{hybrid scatterplot}, it is required to analyze multiple traits simultaneously (\textbf{R2}).
\changjian{We first considered both rectangular bar charts and circular bar charts for encoding multiple traits.
However, the glyphs become invisible when the values of the traits are very small.}
Therefore, following the work of \cite{li2021visual}, we use a glyph with multiple concentric circles (\eg, $\vcenter{\hbox{\includegraphics[height=1.5\fontcharht\font`\B]{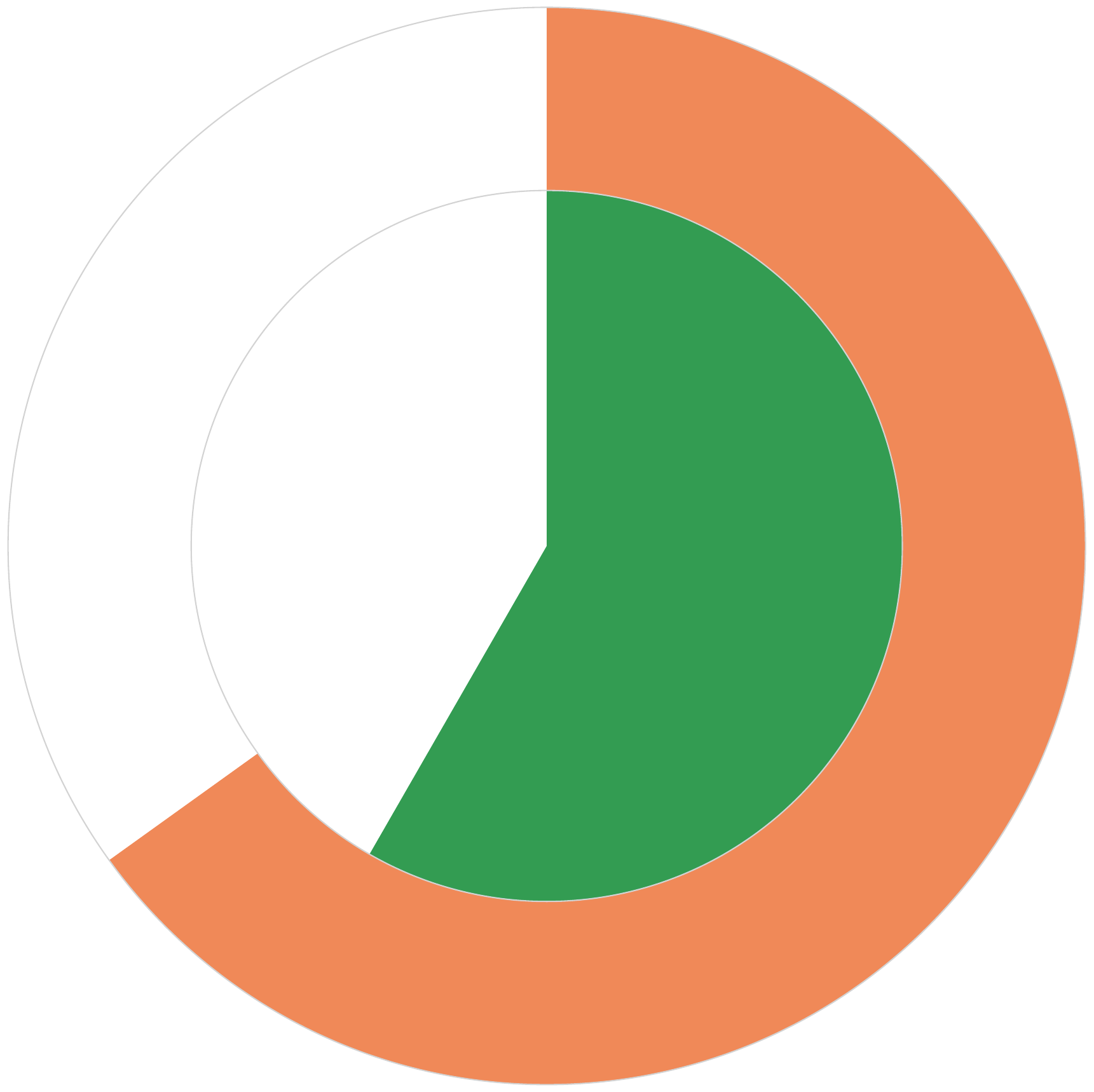}}}$) to represent a hybrid.
The colors of the circles correspond to different trait types, while the fill within the circles indicates the values of these traits.
\changjian{To facilitate better exploration when the number of encoded traits is large, we introduced a filter.
Users can click the legends on the top of the hybrid scatterplot to filter the traits of interest and hide the others.}
In the \textit{gene scatterplot}, each dot represents a gene.
In both scatterplots, the size of the glyphs reflects the associated weights, with the calculation method described in the next section.


\myparagraph{Interactions}.
Within the scatterplots, users modify the positions of the hybrids/genes to explore the relationships between them.  
The available interactions include: 1) selection ($\vcenter{\hbox{\includegraphics[height=1.5\fontcharht\font`\B]{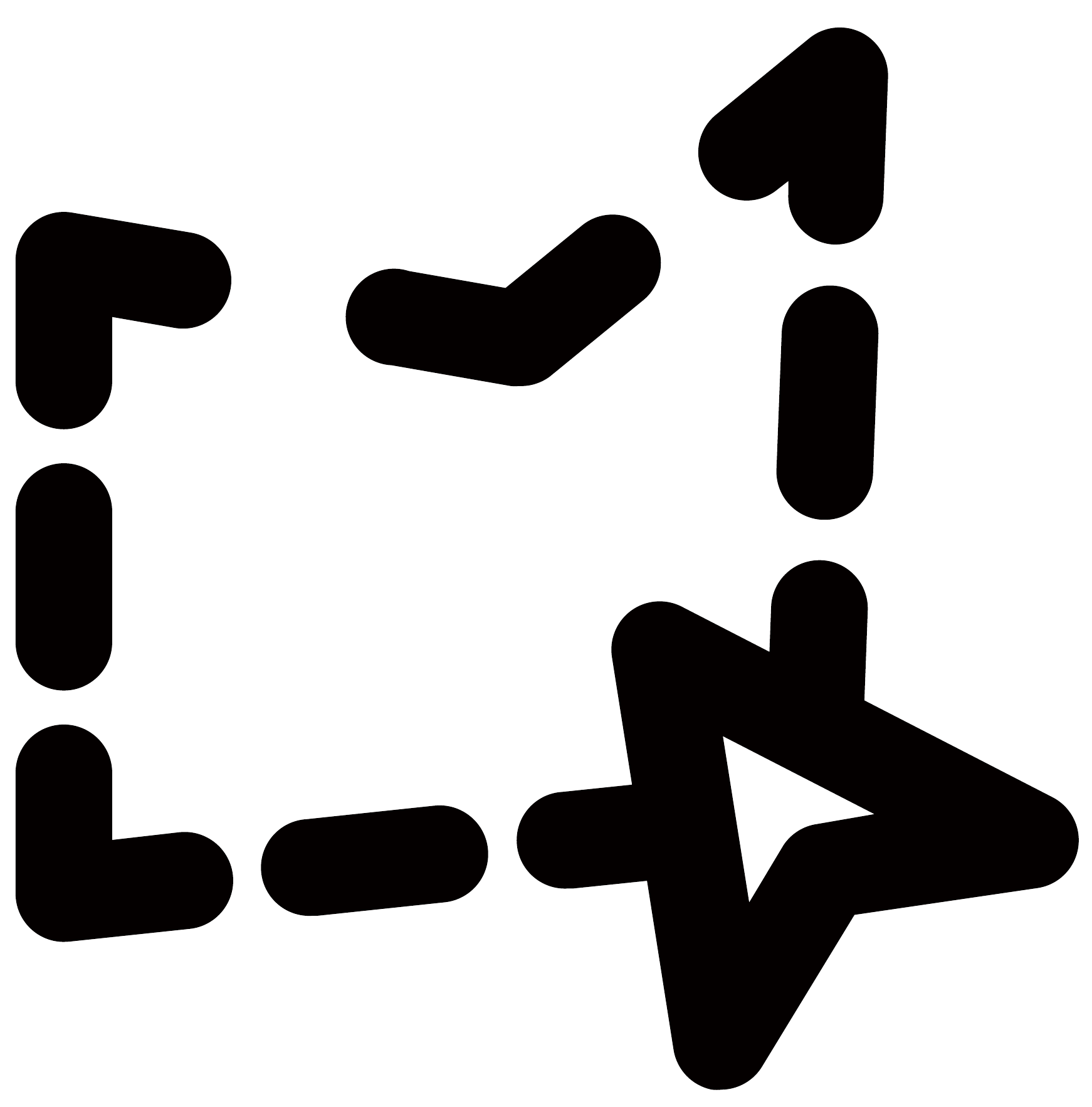}}}$): selecting specific hybrids/genes for further actions; 
2) moving ($\vcenter{\hbox{\includegraphics[height=1.5\fontcharht\font`\B]{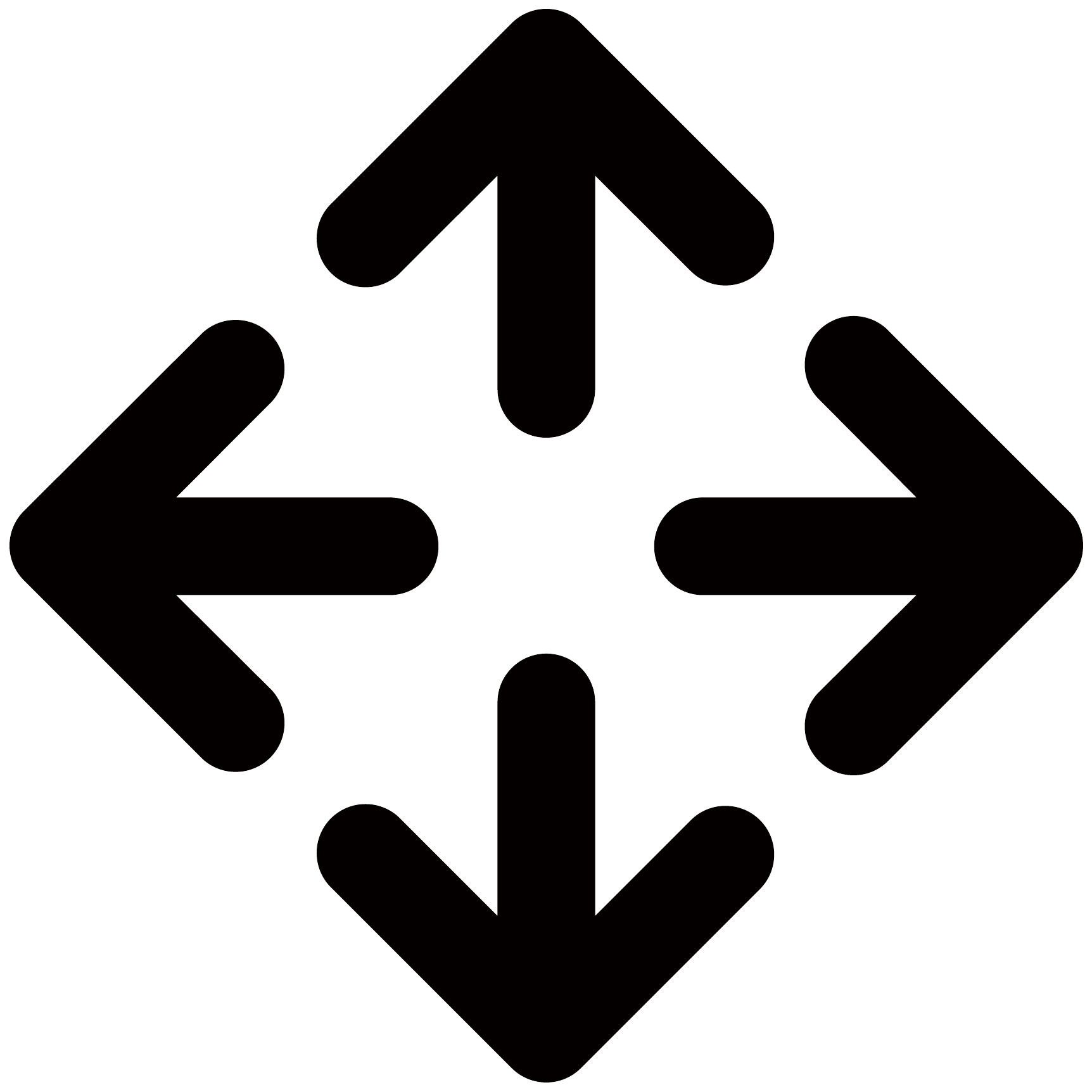}}}$): moving selected hybrids/genes to other areas to indicate changes in their similarities with the others;
3) expanding/contracting ($\vcenter{\hbox{\includegraphics[height=1.5\fontcharht\font`\B]{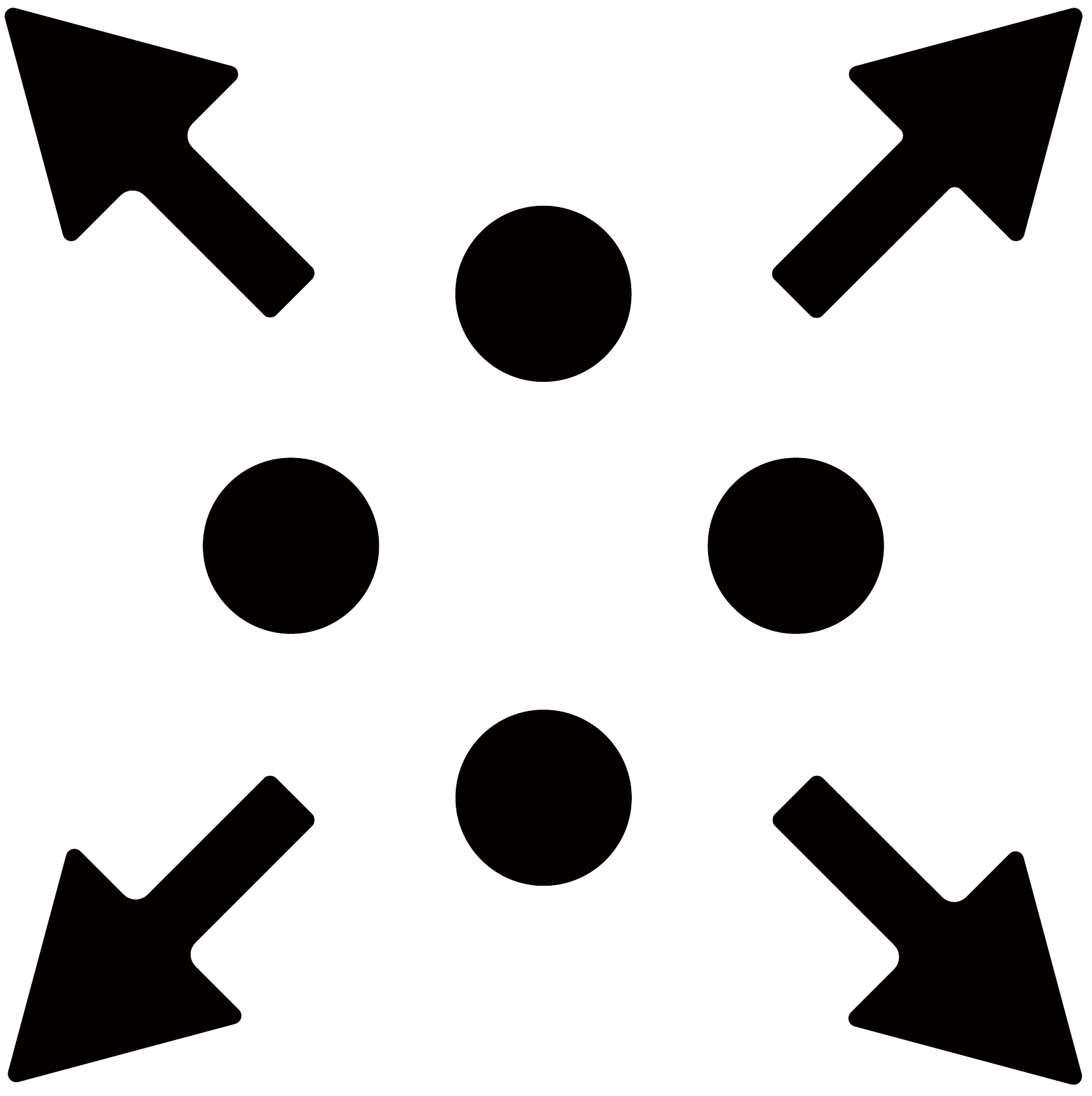}}}$ / $\vcenter{\hbox{\includegraphics[height=1.5\fontcharht\font`\B]{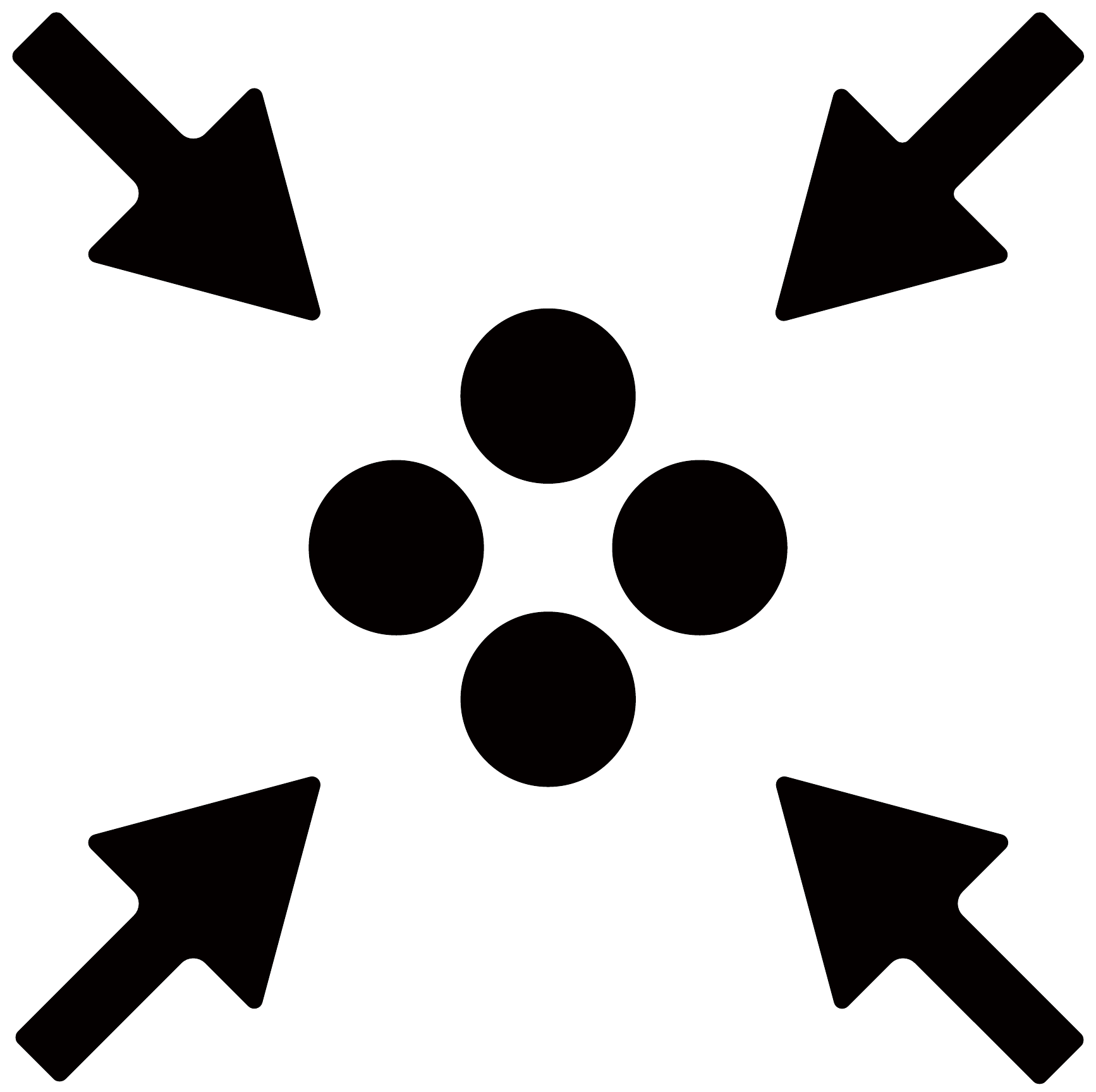}}}$): scaling the selected hybrids/genes to indicate changes in their internal similarities.

When one scatterplot is modified, 
the other scatterplot and the weights of the associated data will be updated accordingly.
The updates to the scatterplot are described in the previous section.
Regarding the calculation of weights, we take modifying $S$ to $S'$ as an example.
Since the updated $X'$ can be obtained by a backward pass of the network, the weights (represented as a diagonal matrix) can be obtained by:
\begin{equation}
    \begin{aligned}
        W = \underset{W}{\arg\min} \| XW-X'\|^2.
    \end{aligned}
\end{equation}
This optimization problem has a closed-form solution given by: $W = {\mathrm{diag}\left (X^T X' \right )} / {\mathrm{diag}\left (X^T X \right )}.$
Here, $\mathrm{diag(\cdot)}$ sets all off-diagonal elements to zero, leaving only the diagonal elements.


\subsection{Hybrid Visualization}
To help breeders select desired hybrids, we developed a hybrid recommendation method to recommend hybrid candidates.
Based on the recommendation, a hybrid visualization (Fig.~\ref{fig:system}(b)) is provided to help analyze the recommended hybrids for selection.

\myparagraph{Hybrid recommendation}.
Given $N$ paternal lines $R=\{r_1, r_2,...,r_N\}$ and $M$ maternal lines $T=\{t_1, t_2,...,t_M\}$, the hybrid recommendation is to select several representative hybrids while minimizing the number of selected hybrids to reduce the workload for field cultivation: 
\begin{equation}
\begin{aligned}
\label{eq:subset-selection}
& \min_{U}  & & \sum_{j}\min_{h_i \in U} D(h_i, h_j) + \gamma \lvert U \rvert, \\
& \mathrm{s.t.} & & D(r_{h_i}, t_{h_i}) > \epsilon, \forall h_i \in U, \\
& & & \mathrm{score}(h_i) > \beta, \forall h_i \in U, \\
& & & h_i \notin H, \forall h_i \in U.
\end{aligned}
\end{equation}
Here, the first term ensures better representativeness by minimizing the sum of the minimum distances between each selected hybrid and the unselected ones.
The second term favors the selection of a small number of hybrids to reduce the workload for field cultivation.
$\gamma$ is the weight to balance these terms and can be set as $\max_{ij} D(h_i,h_j)/K$ to select around $K$ hybrids~\cite{elhamifar2015dissimilarity}.
The first constraint ensures high genomic distances between the paternal and maternal lines of the recommended hybrids.
The second constraint ensures desired traits based on the genomic prediction results. 
The third constraint ensures that the recommended hybrids have not been cultivated.
$U$ is the set of recommended hybrids.
$D(\cdot, \cdot)$ is the genomic distance between two lines, which is measured by the total number of mismatched genes~\cite{lv2020resequencing, chen2022towards}.
$\epsilon$ is the threshold for the genomic distances, which is set as the mean of the genomic distances of the cultivated hybrids in our implementation, and it can be adjusted by the breeders. 
$\mathrm{score}(\cdot)$ is a scoring function that measures the goodness of the traits based on the genomic prediction results.
Its definition depends on the needs of breeders.
For example, if breeders are interested in yields, $\mathrm{score}(\cdot)$ can be set to the predicted yields. 
Alternatively, if breeders are interested in both yields and high length/width ratios, $\mathrm{score}(\cdot)$ can be the sum of yields and length/width ratios.
$\beta$ is the threshold for the score function and can be adjusted by the breeders.
$H$ are the hybrids that have been crossbred and cultivated.
Since optimizing Eq.~(\ref{eq:subset-selection}) is NP-hard, we use the Alternating Direction Method of Multipliers~\cite{elhamifar2015dissimilarity}, an approximate algorithm to solve it. 

\myparagraph{Visual encoding}.
\changjian{In the hybrid visualization, it is important for breeders to simultaneously explore the parental and maternal lines and the hybrid, which naturally form a bipartite graph.
A previous study demonstrated that a node-link diagram is particularly effective in visually highlighting all of them as individual objects~\cite{alsallakh2016state}. 
Therefore, we employ a node-link diagram.}
As shown in Fig.~\ref{fig:system}(b), each dot represents a parental line (left) or a material line (right).
A link between a parental line and a material line represents the associated hybrid.
Users can click $\vcenter{\hbox{\includegraphics[height=1.5\fontcharht\font`\B]{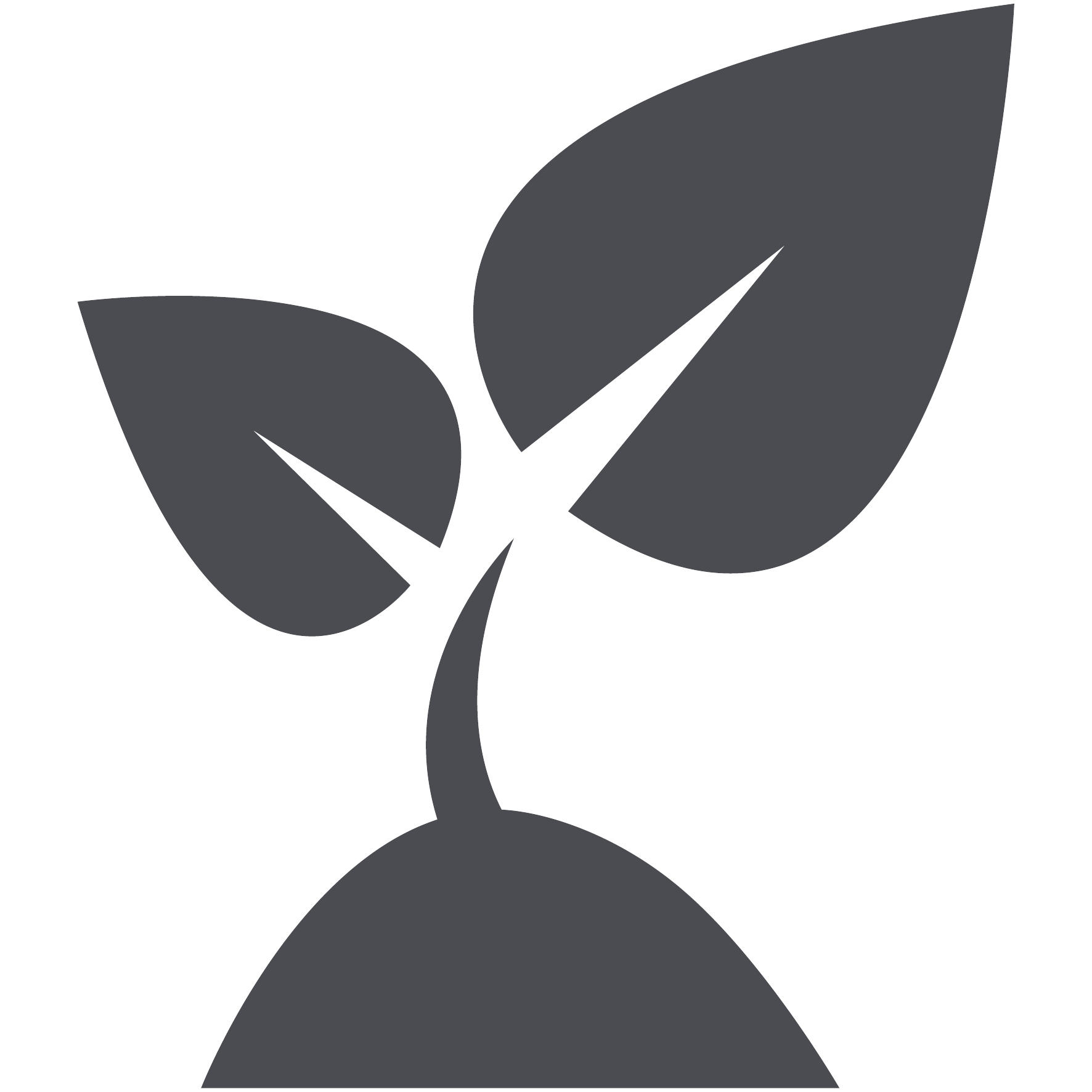}}}$ to recommend several hybrids, which are highlighted in red.

In hybrid breeding, the similarities between parental and maternal lines are important considerations for breeders.
For instance, crossing two similar parental lines with a material line typically results in similar hybrids. 
In such cases, it is unnecessary to cultivate both hybrids.
To facilitate such exploration, similar parental/material lines are expected to be placed together~\cite{chen2020oodanalyzer, liu2019interative}.
Therefore, we follow the work of Mishra~\etal~\cite{mishra2025promptaid} to project parental/material lines with one-dimensional t-SNE.
Edge bundling is utilized to improve readability~\cite{holten2009force, chen2021interactive}. 

\subsection{Gene Visualization}
The gene visualization displays genes in the context of neighbors on chromosomes, which helps analyze similar regulatory genes simultaneously.
As shown in Fig.~\ref{fig:system}(c), the 12 $\vcenter{\hbox{\includegraphics[height=1.5\fontcharht\font`\B]{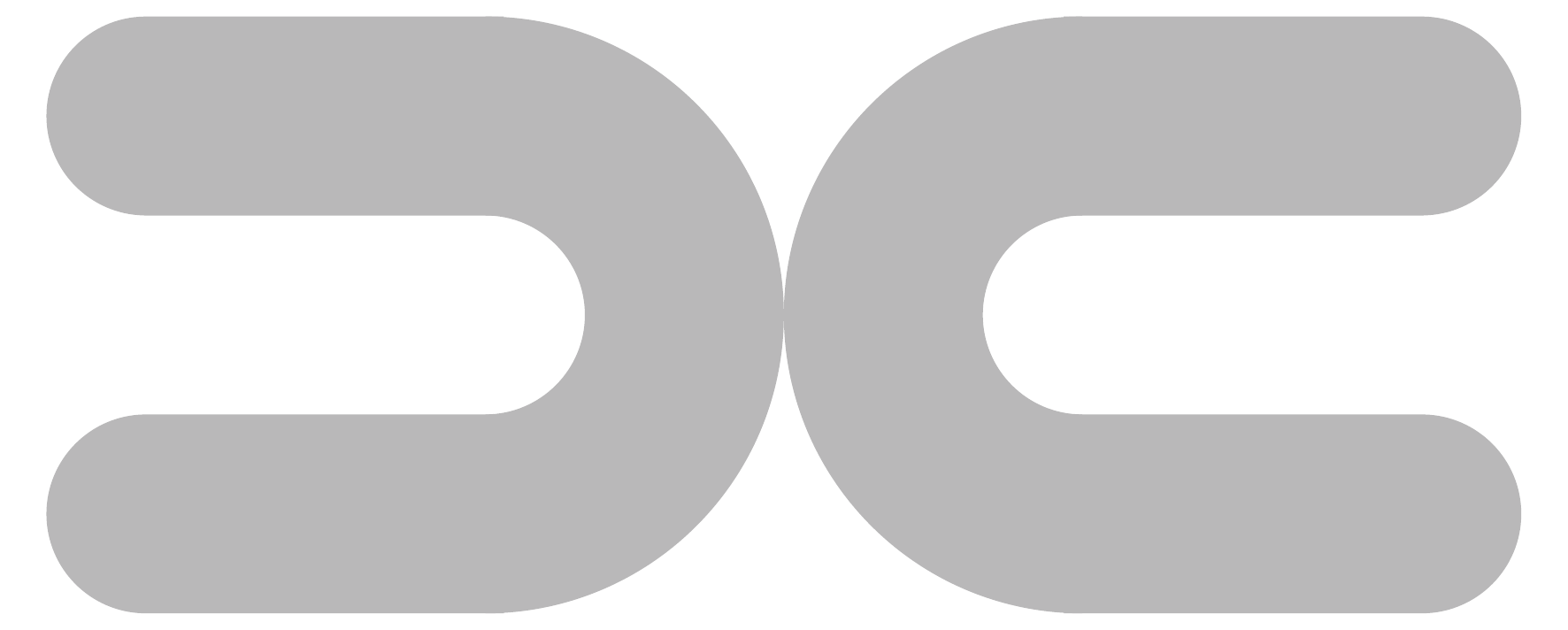}}}$ on the top represents the 12 chromosomes of the rice.
Each row below represents a hybrid, which uses bar charts to represent genes distributed along the chromosomes in order.
Each bar represents a gene.
The height of the bars represents the weights of the genes.
The colors encode genotypes of genes: homo reference, homo alternative, and hetero (\eg, Fig.~\ref{fig:background}(a)).
If users select some genes of interest in the dual analysis visualization, only the associated bars will be displayed.

In the gene visualization, users can select a gene, and a boxplot will be presented (\eg, Fig.~\ref{fig:system}D).
The boxplot presents how the trait values are distributed among different genotypes of the selected gene.
A clear difference in trait values between genotypes helps confirm whether the selected gene controls the traits.

\section{Evaluation}
\label{evaluation}
To demonstrate the effectiveness of the proposed parametric method for dual projection and {\sys} for facilitating hybrid rice breeding, we conducted a dual projection experiment and a case study.

\subsection{Dual Projection Experiment}
\label{subsec:dpe}

\myparagraph{Datasets}.
We conducted the experiments on both image and rice genomic datasets.
The image datasets include MNIST~\cite{lecun1998mnist} and CIFAR-10~\cite{krizhevsky2009cifar-10}. 
The MNIST dataset contains 60,000 training images of handwritten digits ranging from zero to nine.
We randomly sampled 600 images for each class.
The CIFAR-10 dataset contains 60,000 images from ten classes.
We randomly sampled 1,000 images from each class.
The rice genomic dataset consists of 714 rice hybrids and their associated genes, which are collected from the Hunan Hybrid Rice Research Center and publicly-available data.

\myparagraph{Experimental settings}.
We compared the proposed parametric dual projection method with two types of baselines: single projection and dual projection methods.
The single projection methods include t-SNE and PCA~\cite{xia2024parallel, wang2023hetvis}.
The dual projection method includes only SIRIUS.
For the MNIST dataset, similar to the work of LeCun~\cite{lecun1998mnist}, we flatten their pixel values to 784-dimensional vectors as features.
For the CIFAR-10 dataset, we utilized the CLIP model~\cite{radford2021clip} to extract their features. 
For the rice genomic dataset, we use the data processing method described in Sec.~\ref{sec:background} to obtain their features.
\changjian{For our method, four invertible layers are used in the network. 
More details about the network can be found in the supplemental material.}

\myparagraph{Evaluation measures}.
We evaluated the results using the trustworthiness and continuity metrics to measure neighborhood preservation in the projection. 
To demonstrate that our method can meet the interactive requirements, we also compare the computational time of SIRIUS and our method in updating the projection results during dual analysis.

\underline{\emph{{\normalsize Trustworthiness}}}. It measures how well the k-nearest neighbors (kNNs) of a point in the embedding space reflect its true neighbors in the high-dimensional space.
Similar to MFM~\cite{ye2025modalchorus}, we evaluated trustworthiness using k=30 nearest neighbors (\ie, T(30)) to balance effectiveness and computational efficiency.

\underline{{\emph{\normalsize Continuity}}}. It measures how well the kNNs of a point in the high-dimensional space are preserved in the embedding space.  
Similar to Trustworthiness, only 30-nearest neighbors are considered (\ie, C(30)).


\myparagraph{Quantitative projection results}.
The quantitative projection results are shown in Tab.~\ref{table-experiments}.
Our method performs better than the dual projection method SIRIUS across all three datasets.
Additionally, compared to the single projection methods, our method also shows competitive performance. 
For the image datasets, our method keeps the performance gap with the best single method within 2\%. 
For the genomic dataset, our method achieves the best performance.
This validates the effectiveness of our method for preserving neighborhood relationships.

\myparagraph{Qualitative projection results}.
We also visually compare SIRIUS and our method on all three datasets.
Since the comparison in all three datasets leads to similar conclusions, we only show the comparison in the MNIST dataset (Fig.~\ref{fig:projection}) in this manuscript.
The comparison in the other datasets can be found in the supplemental material.
As we can see, for SIRIUS, the classes are cluttered together, which makes visual exploration challenging.
In contrast, our method shows a clear class separation, demonstrating the effectiveness in preserving neighbors.


\myparagraph{Running time in updating}.
To demonstrate that our method can meet the interactive requirements, we compare the running time of SIRIUS and our method for updating the projection results during dual analysis.
The experiment is conducted on a server with an Intel Xeon Silver 4214 CPU and an NVIDIA RTX 2080Ti GPU.
As shown in Tab.~\ref{table-experiments}, our method significantly reduces the running time in updating compared with SIRIUS.
In addition, our method updates in 0.01 seconds, which meets the interactive requirements.

\begin{figure}[t]
    \centering
    \includegraphics[width=\linewidth]{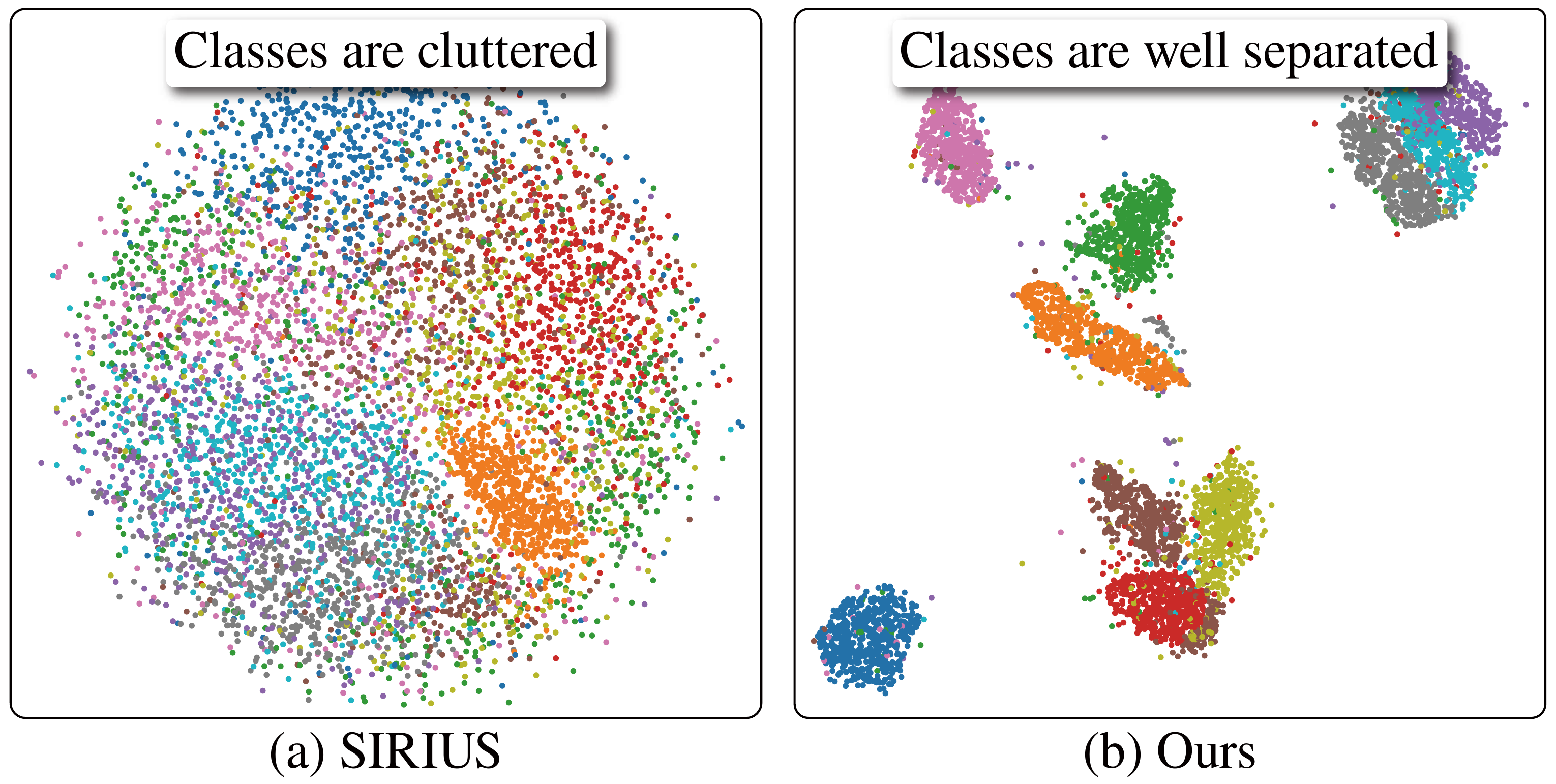}
    \caption{Visual comparison between SIRIUS and our method.}
    \label{fig:projection}
\end{figure}

\subsection{Case Study}
The case study was conducted with B1, one of the breeders involved in the requirement analysis, to demonstrate the effectiveness of {\sys} in facilitating hybrid rice breeding.
B1 aimed to develop a new hybrid of Indica rice in Hunan Province, China.
This hybrid is expected to have a larger length/width ratio to improve the eating quality while also delivering high yields to ensure economic value.
To develop such a hybrid, B1 used {\sys} to analyze the hybrids that have been cultivated in Hunan Province.
In the case study, to allow B1 to focus more on analytical tasks, we used the pair analytics protocol, where we handled the navigation of the tool~\cite{arias2011pair, chen2025human}.



\begin{figure}[t]
    \centering
    \includegraphics[width=\linewidth]{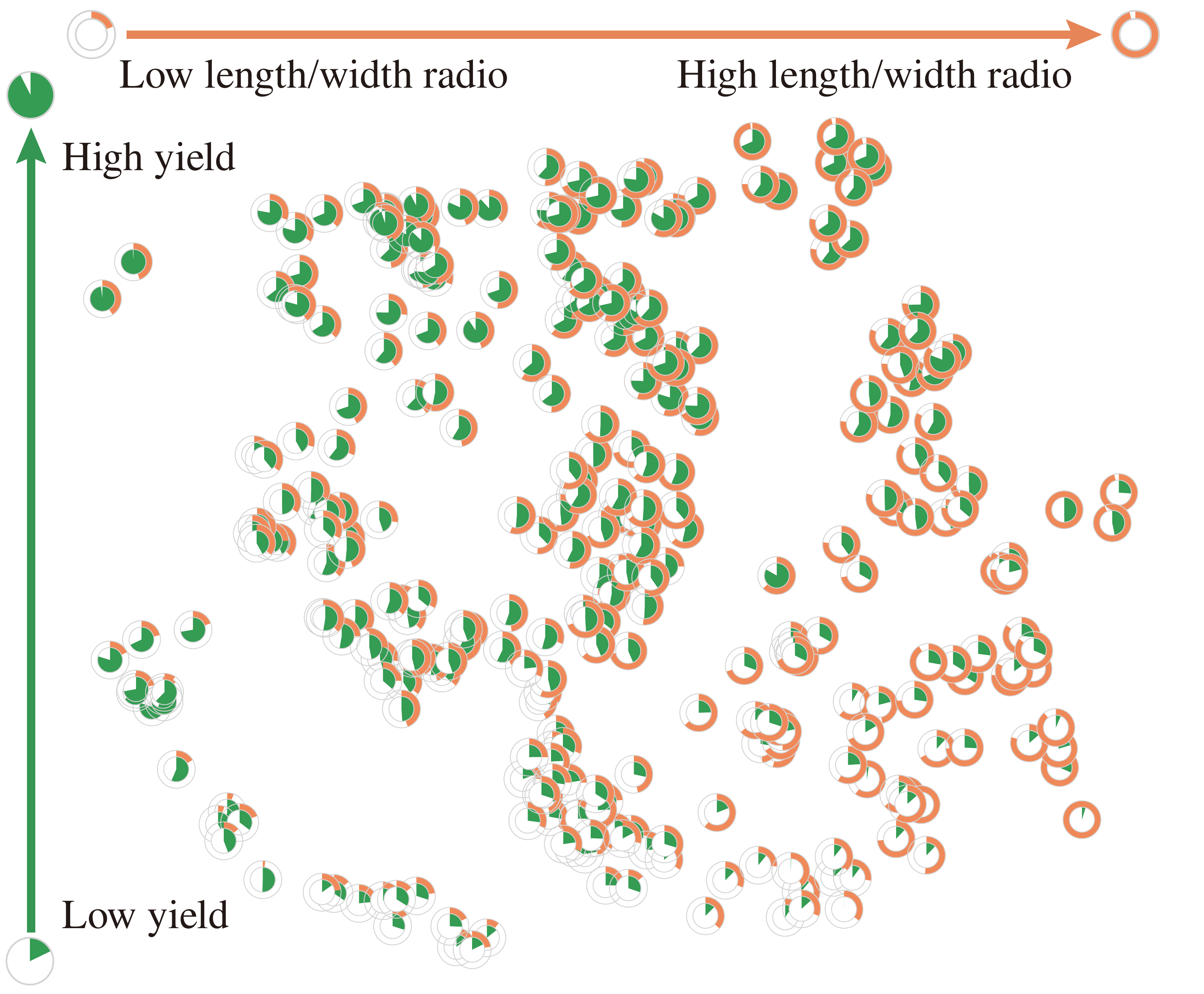}
    \caption{The initial hybrid scatterplot.}
    \label{fig:case-initial}
\end{figure}

\begin{figure}[b]
    \centering
    \includegraphics[width=\linewidth]{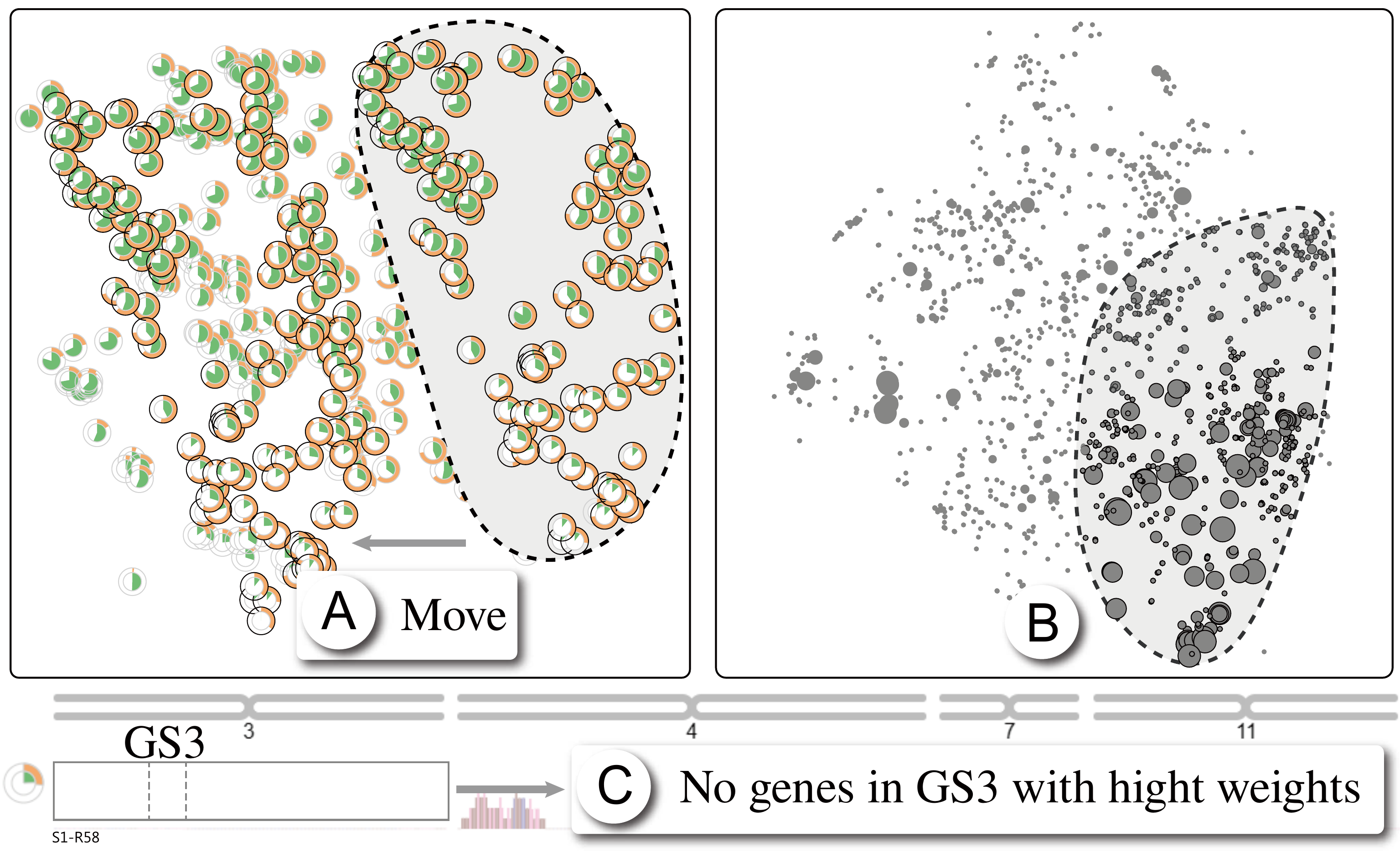}
    \caption{Dual analysis for length/width ratios under all hybrids.}
    \label{fig:case1-step2}
\end{figure}

\begin{figure*}[b]
    \centering
    \includegraphics[width=\linewidth]{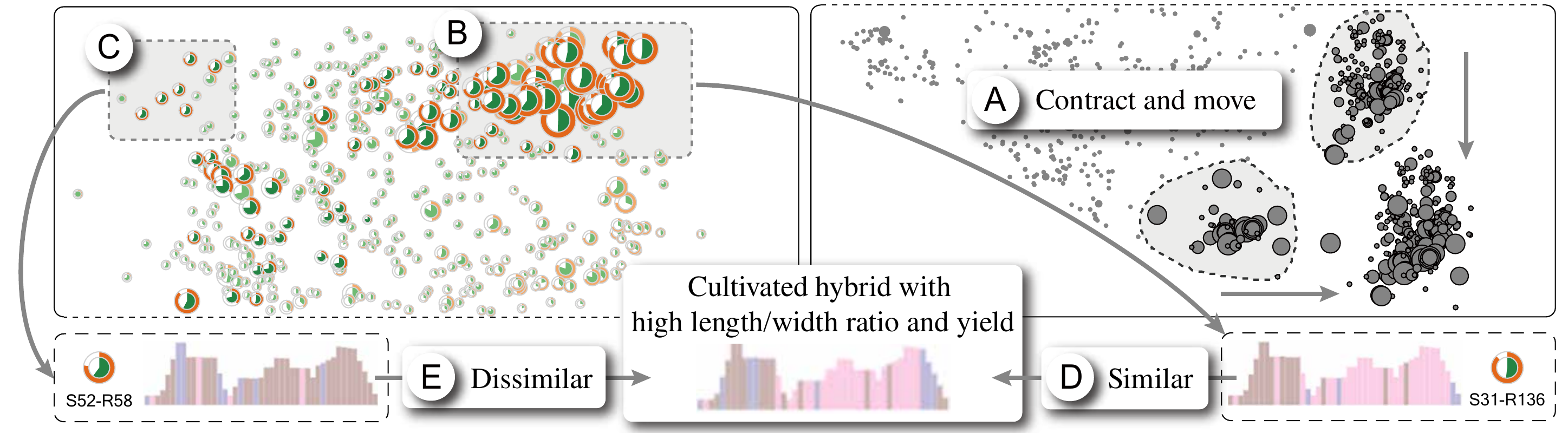}
    \caption{Modifying the gene scatterplot and updating the hybrid scatterplot.}
    \label{fig:case2}
\end{figure*}

\subsubsection{Regulatory Gene Identification}

\myparagraph{Preliminary}.
To begin with, B1 first collected a training dataset of 714 Indica rice hybrids that have been cultivated in Hunan Province and their traits.
It is important to note that collecting the traits of hybrids is time-consuming and challenging. 
Each hybrid usually needs to be cultivated with 200 to 300 plants to reduce randomness, which usually takes about five months before maturation.
Additionally, measuring the traits of the hybrids after maturation usually takes a breeder approximately two months.
Therefore, despite the training dataset only containing 714 hybrids, it is still \textbf{the largest Indica rice dataset}. 
Since B1 mainly focused on the length/width ratio and the yield, he used GWAS~\cite{huang2010genome}, a widely used correlation analysis method, to eliminate genes that had minimal correlation with these two types of traits.

Before the case study, B1 also surveyed the regulatory genes for the length/width ratio and yield in the literature.
The reported regulatory genes only include GS3 in chromosome 3 and GW7 in chromosome 7~\cite{huang2010genome}.
In contrast, the yield is regulated by hundreds of genes~\cite{huang2010genome}, making the analysis of its regulatory genes too complex.
Therefore, B1 mainly focuses on analyzing the regulatory genes of length/width ratios and ensuring yield based on the genomic prediction model.

\myparagraph{Overview}.
B1 began his analysis from the hybrid scatterplot in the dual analysis visualization (Fig.~\ref{fig:case-initial}).
In the hybrid scatterplot, the inner circles ($\vcenter{\hbox{\includegraphics[height=1.5\fontcharht\font`\B]{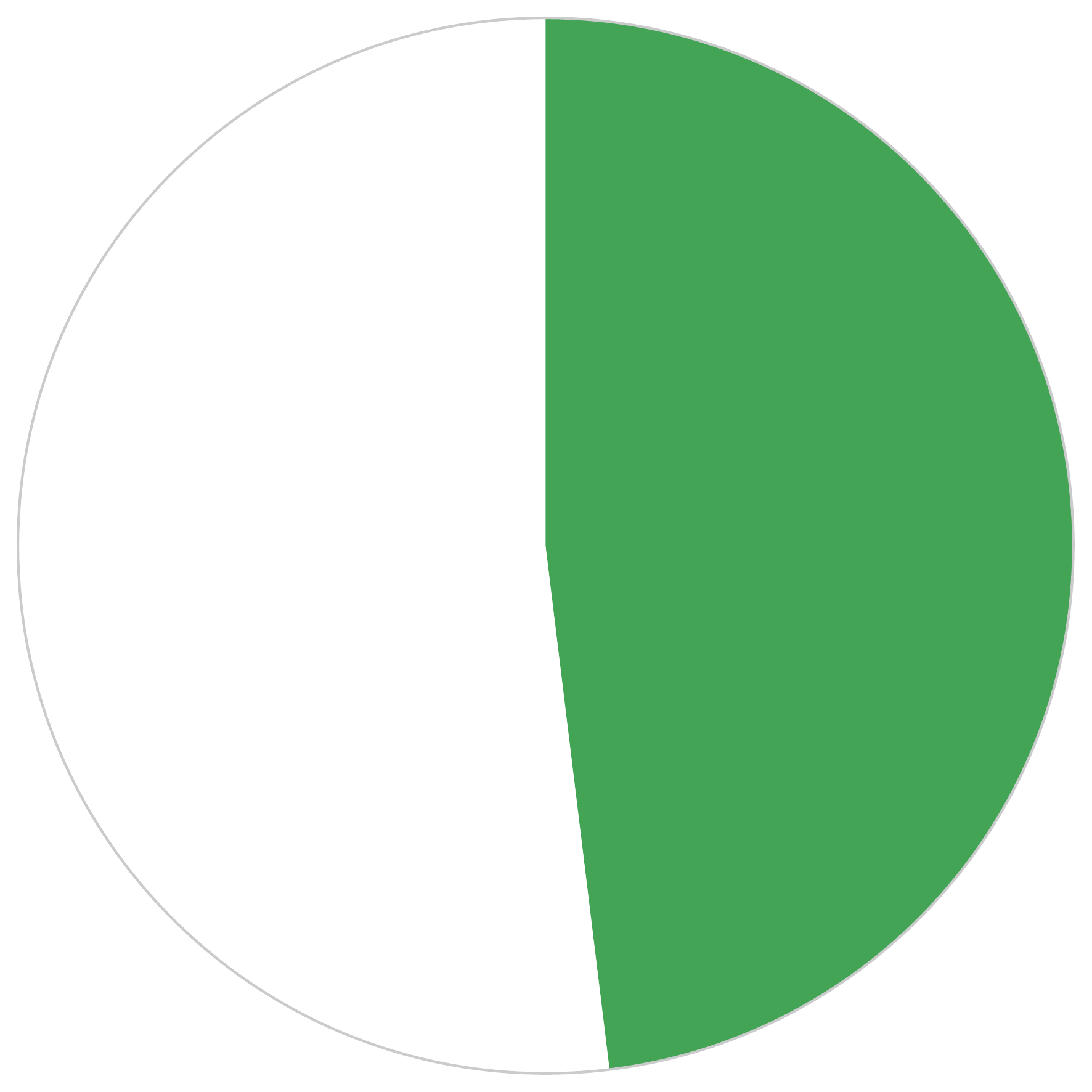}}}$) represent the yields, and the outer circles ($\vcenter{\hbox{\includegraphics[height=1.5\fontcharht\font`\B]{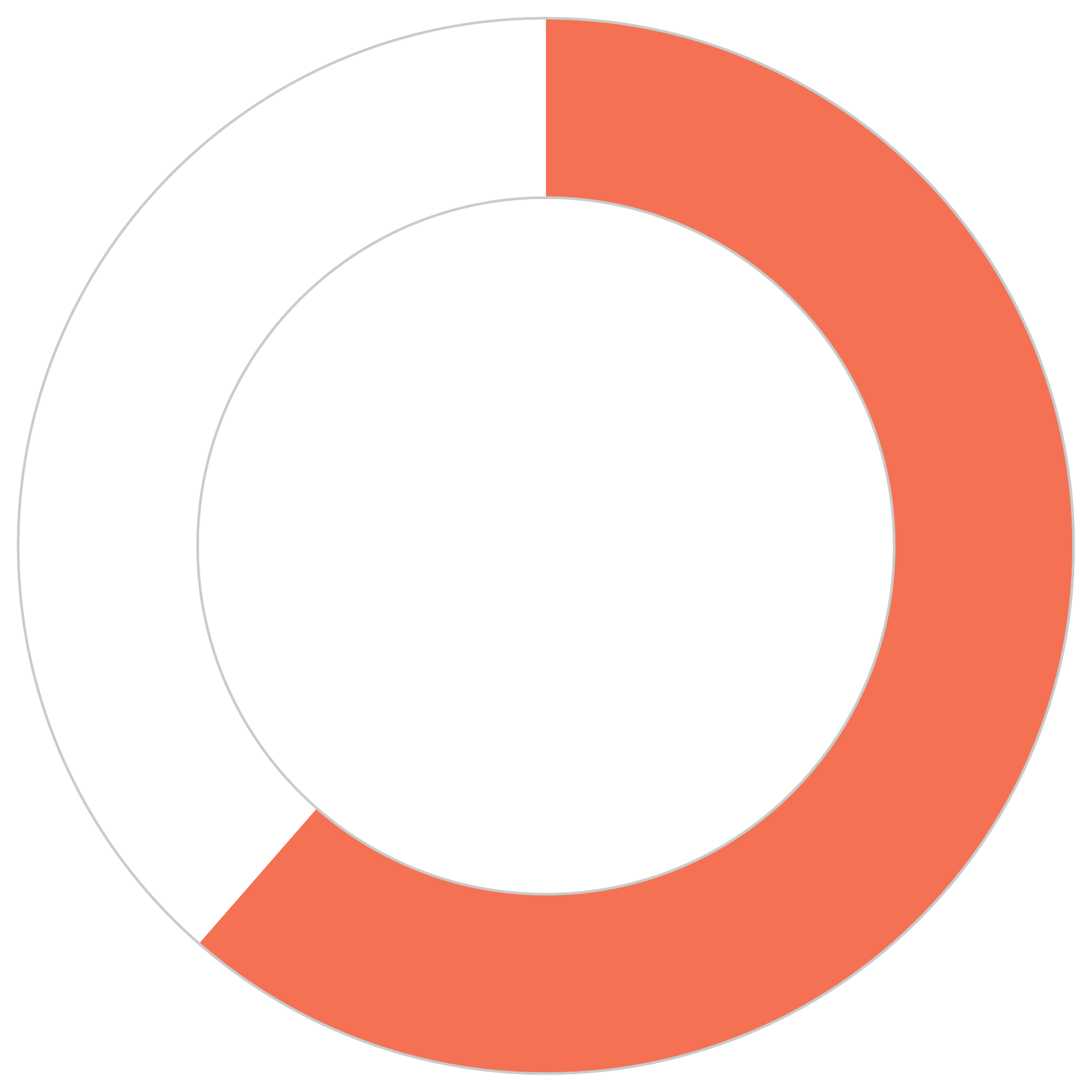}}}$) represent the length/width ratios.
B1 noticed that the fillings of the inner circles increased from bottom to top, indicating that the yields increased in this direction; and the fillings of the outer circles increased from left to right, indicating that the length/width ratios increased in this direction.
Since B1 wanted to develop hybrids with high yields and high length/width ratios, he was especially interested in how the genes regulated the high length/width ratios given high yields.
Therefore, he moved the hybrids in the top-right region (high yields, high length/width ratios) to the top-left region (high yields, low length/width ratios) (Fig.~\ref{fig:system}A).
After that, the gene scatterplot was updated with the proposed parametric dual projection method.

\myparagraph{Identifying regulatory genes regulating length/width ratios when yields are high}.
In the updated gene scatterplot, B1 found that some genes increased in size, indicating higher weights, with the majority gathering in the region B (Fig.~\ref{fig:system}B).
This change in weights reflected B1's modifications in the hybrid scatterplot, suggesting that these genes regulated the length/width ratios when the yields were high.
To further analyze how the genes regulated the length/width ratios, he selected these genes in region B and moved on to gene visualization.

In the gene visualization, B1 found that these genes are mainly distributed on chromosomes 4 and 7 (Figs.~\ref{fig:system}C and \ref{fig:system}E).
B1 first analyzed chromosome 7, where the reported regulatory gene GW7 was located.
In chromosome 7, most of the genes with higher weights are concentrated on a segment (Fig.~\ref{fig:system}C).
By checking this segment, B1 found that GW7 was located in it.
To further verify that these genes regulated the length/width ratio, 
B1 selected one of them and checked how the length/width ratios were distributed among the different genotypes of the selected gene.
As shown in Fig.~\ref{fig:system}D, there was a clear correlation between the selected gene and the length/width ratio, confirming that this gene regulated the length/width ratio.
It demonstrated that our method could help identify regulatory genes that had been reported.
B1 also explained why genes around GW7 also have higher weights: ``Due to factors such as linkage disequilibrium and genomic structure~\cite{flint2003structure}, genes regulating the same trait are usually located at neighboring genomic loci on chromosomes.''

B1 continued to explore chromosome 4, which has not been reported to have regulatory genes related to the length/width ratio.
\begin{wrapfigure}[9]{r}{0.23\textwidth}
 \vspace{-12pt}
\includegraphics[width=0.23\textwidth]{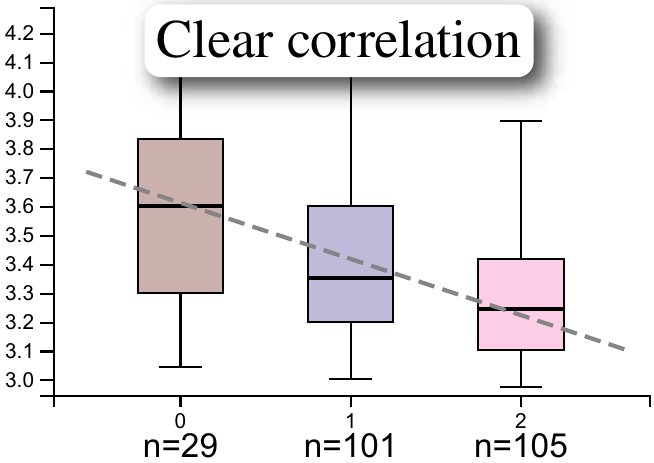}
\end{wrapfigure}
The genes with higher weights concentrated on two segments (Fig.~\ref{fig:system}E).
By checking the distribution of the length/width ratio values across the different genotypes of these genes, B1 also found clear correlations (\eg, the figure on the right). 
B1 explained that discovering unreported regulatory genes was common because the effects of genes could differ in different rice populations.
In their daily work, they also often utilized unreported regulatory genes they discovered for hybrid breeding.

\myparagraph{Identifying regulatory genes with a little mutation}.
After the analysis of chromosomes 4 and 7, B1 was interested in why there were no genes with larger weights in chromosome 3, where the regulatory gene GS3 was located.
He suspected that GS3 might be effective when the yield is medium or low but not when the yield is high.
To verify this, he moved all the hybrids with higher length/width ratios from the right to the left (Fig.~\ref{fig:case1-step2}A).
This helped him identify genes that regulated length/width ratios without restriction to high yield. 
Upon update, B1 observed that genes with higher weights were still positioned in the lower right corner (Fig.~\ref{fig:case1-step2}B).
He selected these genes for further analysis in gene visualization and discovered that GS3 was still absent (Fig.~\ref{fig:case1-step2}C).

This result surprised B1 because GS3 is a well-known regulatory gene for length/width ratios.
To investigate further, he selected GS3 and examined the distribution of length/width ratios in the boxplot.
He found that the majority of hybrids (625) were homo alternative in this gene, and there are no homo reference genes (the figure on the right).
This extreme imbalance in the data caused the fake patterns that GS3 did not regulate the length/width ratio.
\begin{wrapfigure}[9]{r}{0.23\textwidth}
 \vspace{-12pt}
\includegraphics[width=0.23\textwidth]{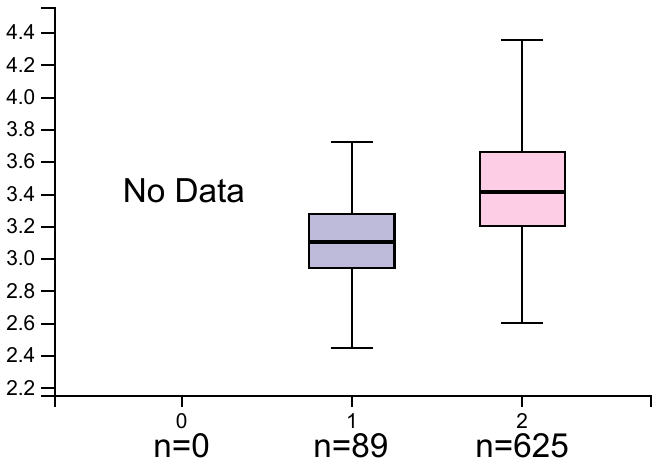}
\end{wrapfigure}
However, B1 remained uncertain about why there was a little mutation in this gene. 
As a result, he consulted B3, who had several years of experience with regulatory genes related to the length/width ratio.
B3 explained, ``homo alternative genotypes usually exist in Indica rice, and the homo reference genotypes usually exist in Japonica rice. 
However, due to the large reproductive isolation between these two types of rice, their hybrid breeding has been unsuccessful in history. Therefore, the homo reference genotypes in Japonica rice are not widely introduced to Indica rice.''
Nevertheless, B3 pointed out that GS3 should also be considered during hybrid breeding because homo reference genotypes in GS3 usually led low length/width ratios.





Afterward, B1 was satisfied with the identified regulatory genes and proceeded to hybrid selection.

\subsubsection{Rice Hybrid Selection}

\myparagraph{Recommending hybrids}.
After identifying the regulatory genes for length/width ratios, B1 wanted to use them to select hybrids with desired traits.
To begin with, he first used the recommendation function of our tool to recommend 150 hybrids.
These recommended hybrids are highlighted in lighter colors in the hybrid scatterplot, and most of them have high length/width ratios and yields (Fig.~\ref{fig:case2}).
However, B1 was not sure whether the recommended hybrids were consistent with the identified regulatory genes.
Therefore, B1 decided to use the dual analysis visualization to analyze these recommended hybrids.
B1 selected the regulatory genes identified in the previous step and contracted and moved them away from the other genes (Fig.~\ref{fig:case2}A).

\myparagraph{Analyzing recommended hybrids and selecting the ones with desired traits}.
Upon update, some recommended hybrids became larger (\eg, Fig.~\ref{fig:case2}B), and some became smaller (\eg, Fig.~\ref{fig:case2}C). 
A larger hybrid indicated a strong correlation with the selected regulatory genes, and a smaller one indicated a weak correlation.
To verify this, B1 selected the 24 large hybrids in region B to analyze their genotypes in the identified regulatory genes.
The genotypes of these hybrids are similar to that of the cultivated hybrids with higher length/width ratios and yields (\eg, Fig.~\ref{fig:case2}D).
In addition, B1 selected these hybrids and analyzed them in the hybrid visualization.
He found that the maternal lines of these hybrids were mainly from the first and third clusters of maternal lines (Fig.~\ref{fig:system}F).
The lines connecting these two clusters were more sparse than those connecting the other clusters, indicating that these maternal lines were less crossbred.
Such maternal lines were of greater interest to B1 because their potential had not been fully explored, making them more likely to produce hybrids with high length/width ratios and yields.
Based on these observations, B1 believed that these 24 large hybrids in region B were more likely to have high length/width ratios and yields, and selected them for further cultivation.

In contrast, although some hybrids were predicted by the genomic prediction model to have high length/width ratios and yields, they became smaller upon update (\eg, Fig.~\ref{fig:case2}C).
To figure out the underlying reasons, he selected one of them and checked its genes. 
As shown in Fig.~\ref{fig:case2}E, its genotypes were dissimilar to that of the cultivated hybrids with higher length/width ratios and yield.
Based on these findings, B1 believed these hybrids were not consistent with the identified regulatory genes and had less potential to have the desired traits, even though they had higher predicted length/width ratios and yields.
Therefore, B1 decided not to select them for cultivation.


\subsubsection{Field Experiment}
To validate that our method can reduce the workload of field cultivation, a field experiment was conducted.
This experiment compared {\sys} with the automatic genomic selection method.
The comparison measures how many hybrids are needed to be cultivated in the field to identify the desired hybrid.
Based on B1's experience, a hybrid is considered desired if it has a high yield (larger than 600 kilograms per Chinese acre) and its length/width ratio exceeds the maximum value of the hybrids in the training dataset.
Otherwise, it is undesired.
To identify the desired hybrid, the genomic selection method requires cultivating 126 hybrids, while our method only requires 24, largely reducing the workload of the field cultivation.

It is important to note that such an experiment would take at least a year and a half. 
The first year is needed to produce the seeds of the hybrids, and the following year would require another six months for cultivating hybrids and collecting traits.
Therefore, to reduce the evaluation time and effort, the experiment was conducted based on real field cultivation data but in a simulated way.
The key to this simulation is using the private data from Longping High-Tech Agriculture Co., Ltd. to determine whether a hybrid is desired.
Longping High-Tech Agriculture Co., Ltd. possesses many cultivated hybrids and their traits, but they are not available due to commercial protection.
Therefore, we asked B1 to help with this.
If a hybrid has been cultivated, whether it was desired was based on its traits.
If a hybrid has not been cultivated, it is considered undesired.
Although such a simulated way is not highly rigorous (\eg, some uncultivated hybrids can be the desired ones), 
the large number of hybrids (102) saved by {\sys} can still demonstrate its effectiveness.

\section{Expert Feedback and Discussion}
\label{sec:discuss}

After the case study, we conducted four interviews with B1-B4 to gather feedback. 
Since B2, B3, and B4 were not involved in the case study, we began by providing them with a 30-minute introduction to the case study.
Following the introduction, each interview lasted between 30 and 50 minutes. 
Overall, the expert feedback was positive regarding the usability of {\sys}.
\changjian{From their feedback and our own experience in developing the tool, we also identified several limitations that require further investigation in the future.}


\subsection{Usability}

\myparagraph{Efficient exploration of the relationships between hybrids and genes}.
All the experts liked the dual analysis method, noting that it enhanced the effectiveness and flexibility of analyzing the relationships between hybrids and genes. 
In the past, they were limited to relying on GWAS to analyze the relationships.
``GWAS only reveals linear relationships, and such linear analysis cannot model the complex relationships between hybrids and genes well,'' B1 said.
``This tool offers a better alternative, and I am willing to use it in rice breeding,'' B2 concluded.


\myparagraph{Combining multiple aspects for final hybrid selection}.
B4 mentioned that he liked how this tool integrates multiple sources of information, such as genomic model prediction results and genomic data, to aid in breeding.
He said, "Currently, when I use genomic model prediction results for breeding, I can only rely on tools like Excel for assistance. 
Since these tools are not designed specifically for breeding, I have to process a lot of information in my mind, which wastes a lot of time and effort.
I believe that this tool can largely reduce my workload."

\myparagraph{\pcheng{Generalization to other domains.}}
\pcheng{
B3 noted that, although the system was developed to meet the specific requirements of hybrid rice breeding, the core method, the parametric dual projection, is domain-agnostic. 
``It can be used for exploring the relationships of other data, such as the proteins and traits.'' he said. 
The dual analysis visualization can also be utilized in other domains as long as the visual encodings are re-designed to meet the domain-specific requirements.
}

\subsection{Limitations and Future Work}

\myparagraph{Combining environment factors}.
At present, we assume a stable environment (\eg, we focus on Hunan Province in the case study), where the traits are primarily determined by genomic data. 
Under this assumption, our method performs well.
\changjian{However, all the breeders pointed out that environmental differences between regions were considerable in practice.}
``Even within the same region, environments can vary from year to year.'' \pcheng{B2 said,} ``It would be interesting to investigate how to integrate environmental factors into genomic prediction models and how to effectively reveal the relationships between these environmental factors and traits.''
This involves difficulties such as collecting environmental data and modeling these environmental factors effectively.

\myparagraph{\pcheng{Incremental update of the parametric dual projection method.}}
\pcheng{Currently, when the dataset is fixed or undergoes small changes, the parametric dual projection method works well by fixing the trained parameters. However, when the dataset undergoes large changes (\eg, a 50\% increase in the number of samples), the current parameters cannot fit the changed dataset well and need to be updated accordingly. However, re-training the parametric dual projection method from scratch is time-consuming (\eg, taking more than 12 minutes for the MNIST dataset), which disrupts the analysis process of users. Therefore, it would be interesting to investigate how to incrementally update the parametric dual projection method when the dataset is changed largely.}

\myparagraph{\pcheng{Extensive evaluation.}}
\pcheng{Currently, to allow the breeders to focus more on analytical tasks, we implemented the pair analytics protocol in the case study. However, when the proposed tool is deployed in real-world breeding scenarios, breeders will need to navigate with the tool on their own. Under such circumstances, it remains unexplored how the learning curves behave and whether usability will be compromised. Therefore, we plan to share it with more breeders and collect their feedback to further improve its usability.}

\section{Conclusion}
In this paper, we proposed a visual analysis method to facilitate interactive hybrid rice breeding. 
The key contribution is a parametric dual projection method that enables efficient dual analysis for regulatory gene identification and hybrid selection. 
Based on the dual analysis, a gene visualization and a hybrid visualization are provided to verify the identified regulatory genes and selected hybrids. 
Through quantitative evaluation and a case study, we demonstrated the effectiveness of the parametric dual projection method and the positive outcomes in identifying regulatory genes and desirable hybrids. 
Additionally, positive feedback from the breeders highlights the usefulness of our method in improving the hybrid rice breeding process.

\acknowledgments{
\pcheng{The work is supported by the National Natural Science Foundation of China (Grant Nos. 62225205, 62373141, 62402167), the Science and Technology Program of Changsha (kh2301011), the Science and Technology Innovation Program of Hunan Province (No. 2023ZJ1080), the Hunan Natural Science Foundation under the grant 2025JJ60419, the Major Science and Technology Research Projects of Hunan Province (Grant Nos. 2024QK2010, 2024QK2009).}
}

\bibliographystyle{abbrv-doi-hyperref}

\bibliography{reference}

\begin{thebibliography}{10}

\bibitem{alsallakh2016state}
B.~Alsallakh, L.~Micallef, W.~Aigner, H.~Hauser, S.~Miksch, and P.~Rodgers.
\newblock The state-of-the-art of set visualization.
\newblock {\em Computer Graphics Forum}, 35(1):234--260, 2016. \href{https://doi.org/10.1111/cgf.12722}
{doi: {{%
10\hspace{.1pt}\discretionary{.}{%
}{.}\hspace{.4pt}1111\discretionary{/}{%
}{/}cgf\hspace{.1pt}\discretionary{.}{%
}{.}\hspace{.4pt}12722}}}


\bibitem{arias2011pair}
R.~Arias-Hernandez, L.~T. Kaastra, T.~M. Green, and B.~Fisher.
\newblock Pair analytics: Capturing reasoning processes in collaborative visual analytics.
\newblock In {\em Proceedings of the IEEE International Conference on System Sciences}, pp. 1--10, 2011. \href{https://doi.org/10.1109/HICSS.2011.339}
{doi: {{%
10\hspace{.1pt}\discretionary{.}{%
}{.}\hspace{.4pt}1109\discretionary{/}{%
}{/}HICSS\hspace{.1pt}\discretionary{.}{%
}{.}\hspace{.4pt}2011\hspace{.1pt}\discretionary{.}{%
}{.}\hspace{.4pt}339}}}


\bibitem{artur2019novel}
E.~Artur and R.~Minghim.
\newblock A novel visual approach for enhanced attribute analysis and selection.
\newblock {\em Computers \& Graphics}, 84:160--172, 2019. \href{https://doi.org/10.1016/j.cag.2019.08.015}
{doi: {{%
10\hspace{.1pt}\discretionary{.}{%
}{.}\hspace{.4pt}1016\discretionary{/}{%
}{/}j\hspace{.1pt}\discretionary{.}{%
}{.}\hspace{.4pt}cag\hspace{.1pt}\discretionary{.}{%
}{.}\hspace{.4pt}2019\hspace{.1pt}\discretionary{.}{%
}{.}\hspace{.4pt}08\hspace{.1pt}\discretionary{.}{%
}{.}\hspace{.4pt}015}}}


\bibitem{bard2013practical}
J.~F. Bard.
\newblock {\em Practical bilevel optimization: algorithms and applications}, vol.~30.
\newblock Springer Science \& Business Media, 2013.

\bibitem{blanco2021seqcode}
E.~Blanco, M.~Gonz{\'a}lez-Ram{\'\i}rez, and L.~Di~Croce.
\newblock Productive visualization of high-throughput sequencing data using the seqcode open portable platform.
\newblock {\em Scientific Reports}, 11(1):19545, 2021. \href{https://doi.org/10.1038/s41598-021-98889-7}
{doi: {{%
10\hspace{.1pt}\discretionary{.}{%
}{.}\hspace{.4pt}1038\discretionary{/}{%
}{/}s41598\discretionary{%
}{-}{-}021\discretionary{%
}{-}{-}98889\discretionary{%
}{-}{-}7}}}


\bibitem{chen2024enhancing}
C.~Chen, J.~Chen, W.~Yang, H.~Wang, J.~Knittel, X.~Zhao, S.~Koch, T.~Ertl, and S.~Liu.
\newblock Enhancing single-frame supervision for better temporal action localization.
\newblock {\em IEEE Transactions on Visualization and Computer Graphics}, 30(6):2903--2915, 2024. \href{https://doi.org/10.1109/TVCG.2024.3388521}
{doi: {{%
10\hspace{.1pt}\discretionary{.}{%
}{.}\hspace{.4pt}1109\discretionary{/}{%
}{/}TVCG\hspace{.1pt}\discretionary{.}{%
}{.}\hspace{.4pt}2024\hspace{.1pt}\discretionary{.}{%
}{.}\hspace{.4pt}3388521}}}


\bibitem{chen2025human}
C.~Chen, F.~Lv, Y.~Guan, P.~Wang, S.~Yu, Y.~Zhang, and Z.~Tang.
\newblock Human-guided image generation for expanding small-scale training image datasets.
\newblock {\em IEEE Transactions on Visualization and Computer Graphics}, 31(6):3809--3821, 2025. \href{https://doi.org/10.1109/TVCG.2025.3567053}
{doi: {{%
10\hspace{.1pt}\discretionary{.}{%
}{.}\hspace{.4pt}1109\discretionary{/}{%
}{/}TVCG\hspace{.1pt}\discretionary{.}{%
}{.}\hspace{.4pt}2025\hspace{.1pt}\discretionary{.}{%
}{.}\hspace{.4pt}3567053}}}


\bibitem{chen2021interactive}
C.~Chen, Z.~Wang, J.~Wu, X.~Wang, L.-Z. Guo, Y.-F. Li, and S.~Liu.
\newblock Interactive graph construction for graph-based semi-supervised learning.
\newblock {\em IEEE Transactions on Visualization and Computer Graphics}, 27(9):3701--3716, 2021. \href{https://doi.org/10.1109/TVCG.2021.3084694}
{doi: {{%
10\hspace{.1pt}\discretionary{.}{%
}{.}\hspace{.4pt}1109\discretionary{/}{%
}{/}TVCG\hspace{.1pt}\discretionary{.}{%
}{.}\hspace{.4pt}2021\hspace{.1pt}\discretionary{.}{%
}{.}\hspace{.4pt}3084694}}}


\bibitem{chen2022towards}
C.~Chen, J.~Wu, X.~Wang, S.~Xiang, S.-H. Zhang, Q.~Tang, and S.~Liu.
\newblock Towards better caption supervision for object detection.
\newblock {\em IEEE Transactions on Visualization and Computer Graphics}, 28(4):1941--1954, 2022. \href{https://doi.org/10.1109/TVCG.2021.3138933}
{doi: {{%
10\hspace{.1pt}\discretionary{.}{%
}{.}\hspace{.4pt}1109\discretionary{/}{%
}{/}TVCG\hspace{.1pt}\discretionary{.}{%
}{.}\hspace{.4pt}2021\hspace{.1pt}\discretionary{.}{%
}{.}\hspace{.4pt}3138933}}}


\bibitem{chen2020oodanalyzer}
C.~Chen, J.~Yuan, Y.~Lu, Y.~Liu, H.~Su, S.~Yuan, and S.~Liu.
\newblock {OoDAnalyzer}: Interactive analysis of out-of-distribution samples.
\newblock {\em IEEE Transactions on Visualization and Computer Graphics}, 27(7):3335--3349, 2021. \href{https://doi.org/10.1109/TVCG.2020.2973258}
{doi: {{%
10\hspace{.1pt}\discretionary{.}{%
}{.}\hspace{.4pt}1109\discretionary{/}{%
}{/}TVCG\hspace{.1pt}\discretionary{.}{%
}{.}\hspace{.4pt}2020\hspace{.1pt}\discretionary{.}{%
}{.}\hspace{.4pt}2973258}}}


\bibitem{dennig2023fs}
F.~L. Dennig, M.~Miller, D.~A. Keim, and M.~El-Assady.
\newblock {FS/DS}: A theoretical framework for the dual analysis of feature space and data space.
\newblock {\em IEEE Transactions on Visualization and Computer Graphics}, 2023. \href{https://doi.org/10.1109/TVCG.2023.3288356}
{doi: {{%
10\hspace{.1pt}\discretionary{.}{%
}{.}\hspace{.4pt}1109\discretionary{/}{%
}{/}TVCG\hspace{.1pt}\discretionary{.}{%
}{.}\hspace{.4pt}2023\hspace{.1pt}\discretionary{.}{%
}{.}\hspace{.4pt}3288356}}}


\bibitem{dowling2018sirius}
M.~Dowling, J.~Wenskovitch, J.~Fry, L.~House, and C.~North.
\newblock {SIRIUS}: Dual, symmetric, interactive dimension reductions.
\newblock {\em IEEE Transactions on Visualization and Computer Graphics}, 25(1):172--182, 2018. \href{https://doi.org/10.1109/TVCG.2018.2865047}
{doi: {{%
10\hspace{.1pt}\discretionary{.}{%
}{.}\hspace{.4pt}1109\discretionary{/}{%
}{/}TVCG\hspace{.1pt}\discretionary{.}{%
}{.}\hspace{.4pt}2018\hspace{.1pt}\discretionary{.}{%
}{.}\hspace{.4pt}2865047}}}


\bibitem{elhamifar2015dissimilarity}
E.~Elhamifar, G.~Sapiro, and S.~S. Sastry.
\newblock Dissimilarity-based sparse subset selection.
\newblock {\em IEEE Transactions on Pattern Analysis and Machine Intelligence}, 38(11):2182--2197, 2015. \href{https://doi.org/10.1109/tpami.2015.2511748}
{doi: {{%
10\hspace{.1pt}\discretionary{.}{%
}{.}\hspace{.4pt}1109\discretionary{/}{%
}{/}tpami\hspace{.1pt}\discretionary{.}{%
}{.}\hspace{.4pt}2015\hspace{.1pt}\discretionary{.}{%
}{.}\hspace{.4pt}2511748}}}


\bibitem{endert2011observation}
A.~Endert, C.~Han, D.~Maiti, L.~House, and C.~North.
\newblock Observation-level interaction with statistical models for visual analytics.
\newblock In {\em Proceedings of the IEEE Conference on Visual Analytics Science and Technology}, pp. 121--130, 2011. \href{https://doi.org/10.1109/VAST.2011.6102449}
{doi: {{%
10\hspace{.1pt}\discretionary{.}{%
}{.}\hspace{.4pt}1109\discretionary{/}{%
}{/}VAST\hspace{.1pt}\discretionary{.}{%
}{.}\hspace{.4pt}2011\hspace{.1pt}\discretionary{.}{%
}{.}\hspace{.4pt}6102449}}}


\bibitem{fernstad2013quality}
S.~J. Fernstad, J.~Shaw, and J.~Johansson.
\newblock Quality-based guidance for exploratory dimensionality reduction.
\newblock {\em Information Visualization}, 12(1):44--64, 2013. \href{https://doi.org/10.1177/1473871612460526}
{doi: {{%
10\hspace{.1pt}\discretionary{.}{%
}{.}\hspace{.4pt}1177\discretionary{/}{%
}{/}1473871612460526}}}


\bibitem{flint2003structure}
S.~A. Flint-Garcia, J.~M. Thornsberry, and E.~S. Buckler~IV.
\newblock Structure of linkage disequilibrium in plants.
\newblock {\em Annual Review of Plant Biology}, 54(1):357--374, 2003. \href{https://doi.org/10.1146/annurev.arplant.54.031902.134907}
{doi: {{%
10\hspace{.1pt}\discretionary{.}{%
}{.}\hspace{.4pt}1146\discretionary{/}{%
}{/}annurev\hspace{.1pt}\discretionary{.}{%
}{.}\hspace{.4pt}arplant\hspace{.1pt}\discretionary{.}{%
}{.}\hspace{.4pt}54\hspace{.1pt}\discretionary{.}{%
}{.}\hspace{.4pt}031902\hspace{.1pt}\discretionary{.}{%
}{.}\hspace{.4pt}134907}}}


\bibitem{gansner2005graph}
E.~R. Gansner, Y.~Koren, and S.~North.
\newblock Graph drawing by stress majorization.
\newblock In {\em Graph Drawing: 12th International Symposium, GD 2004, New York, NY, USA, September 29-October 2, 2004, Revised Selected Papers 12}, pp. 239--250, 2005. \href{https://doi.org/10.1007/978-3-540-31843-9_25}
{doi: {{%
10\hspace{.1pt}\discretionary{.}{%
}{.}\hspace{.4pt}1007\discretionary{/}{%
}{/}978\discretionary{%
}{-}{-}3\discretionary{%
}{-}{-}540\discretionary{%
}{-}{-}31843\discretionary{%
}{-}{-}9\_25}}}


\bibitem{garrison2021dimlift}
L.~Garrison, J.~M{\"u}ller, S.~Schreiber, S.~Oeltze-Jafra, H.~Hauser, and S.~Bruckner.
\newblock Dimlift: Interactive hierarchical data exploration through dimensional bundling.
\newblock {\em IEEE Transactions on Visualization and Computer Graphics}, 27(6):2908--2922, 2021. \href{https://doi.org/10.1109/TVCG.2021.3057519}
{doi: {{%
10\hspace{.1pt}\discretionary{.}{%
}{.}\hspace{.4pt}1109\discretionary{/}{%
}{/}TVCG\hspace{.1pt}\discretionary{.}{%
}{.}\hspace{.4pt}2021\hspace{.1pt}\discretionary{.}{%
}{.}\hspace{.4pt}3057519}}}


\bibitem{gu2023structure}
Z.~Gu, J.~Gong, Z.~Zhu, Z.~Li, Q.~Feng, C.~Wang, Y.~Zhao, Q.~Zhan, C.~Zhou, A.~Wang, T.~Huang, L.~Zhang, Q.~Tian, D.~Fan, Y.~Lu, Q.~Zhao, X.~Huang, S.~Yang, and B.~Han.
\newblock Structure and function of rice hybrid genomes reveal genetic basis and optimal performance of heterosis.
\newblock {\em Nature Genetics}, 55(10):1745–1756, 2023. \href{https://doi.org/10.1038/s41588-023-01495-8}
{doi: {{%
10\hspace{.1pt}\discretionary{.}{%
}{.}\hspace{.4pt}1038\discretionary{/}{%
}{/}s41588\discretionary{%
}{-}{-}023\discretionary{%
}{-}{-}01495\discretionary{%
}{-}{-}8}}}


\bibitem{holten2009force}
D.~Holten and J.~J. Van~Wijk.
\newblock Force-directed edge bundling for graph visualization.
\newblock {\em Computer Graphics Forum}, 28(3):983--990, 2009. \href{https://doi.org/10.1111/j.1467-8659.2009.01450.x}
{doi: {{%
10\hspace{.1pt}\discretionary{.}{%
}{.}\hspace{.4pt}1111\discretionary{/}{%
}{/}j\hspace{.1pt}\discretionary{.}{%
}{.}\hspace{.4pt}1467\discretionary{%
}{-}{-}8659\hspace{.1pt}\discretionary{.}{%
}{.}\hspace{.4pt}2009\hspace{.1pt}\discretionary{.}{%
}{.}\hspace{.4pt}01450\hspace{.1pt}\discretionary{.}{%
}{.}\hspace{.4pt}x}}}


\bibitem{huang2010genome}
X.~Huang, X.~Wei, T.~Sang, Q.~Zhao, Q.~Feng, Y.~Zhao, C.~Li, C.~Zhu, T.~Lu, Z.~Zhang, M.~Li, D.~Fan, Y.~Guo, A.~Wang, L.~Wang, L.~Deng, W.~Li, Y.~Lu, Q.~Weng, K.~Liu, T.~Huang, T.~Zhou, Y.~Jing, W.~Li, Z.~Lin, E.~S. Buckler, Q.~Qian, Q.-F. Zhang, J.~Li, and B.~Han.
\newblock Genome-wide association studies of 14 agronomic traits in rice landraces.
\newblock {\em Nature Genetics}, 42(11):961–967, 2010. \href{https://doi.org/10.1038/ng.695}
{doi: {{%
10\hspace{.1pt}\discretionary{.}{%
}{.}\hspace{.4pt}1038\discretionary{/}{%
}{/}ng\hspace{.1pt}\discretionary{.}{%
}{.}\hspace{.4pt}695}}}


\bibitem{jing2021hinet}
J.~Jing, X.~Deng, M.~Xu, J.~Wang, and Z.~Guan.
\newblock {HiNet}: Deep image hiding by invertible network.
\newblock In {\em Proceedings of the IEEE/CVF International Conference on Computer Vision}, pp. 4733--4742, 2021. \href{https://doi.org/10.1109/ICCV48922.2021.00469}
{doi: {{%
10\hspace{.1pt}\discretionary{.}{%
}{.}\hspace{.4pt}1109\discretionary{/}{%
}{/}ICCV48922\hspace{.1pt}\discretionary{.}{%
}{.}\hspace{.4pt}2021\hspace{.1pt}\discretionary{.}{%
}{.}\hspace{.4pt}00469}}}


\bibitem{kawahara2013improvement}
Y.~Kawahara, M.~de~la Bastide, J.~P. Hamilton, H.~Kanamori, W.~R. McCombie, S.~Ouyang, D.~C. Schwartz, T.~Tanaka, J.~Wu, S.~Zhou, et~al.
\newblock Improvement of the oryza sativa nipponbare reference genome using next generation sequence and optical map data.
\newblock {\em Rice}, 6:1--10, 2013. \href{https://doi.org/10.1186/1939-8433-6-4}
{doi: {{%
10\hspace{.1pt}\discretionary{.}{%
}{.}\hspace{.4pt}1186\discretionary{/}{%
}{/}1939\discretionary{%
}{-}{-}8433\discretionary{%
}{-}{-}6\discretionary{%
}{-}{-}4}}}


\bibitem{kerkhoven2004mgv}
R.~Kerkhoven, F.~H. Van~Enckevort, J.~Boekhorst, D.~Molenaar, and R.~J. Siezen.
\newblock Visualization for genomics: the microbial genome viewer.
\newblock {\em Bioinformatics}, 20(11):1812--1814, 2004. \href{https://doi.org/10.1093/bioinformatics/bth159}
{doi: {{%
10\hspace{.1pt}\discretionary{.}{%
}{.}\hspace{.4pt}1093\discretionary{/}{%
}{/}bioinformatics\discretionary{/}{%
}{/}bth159}}}


\bibitem{kingma2018glow}
D.~P. Kingma and P.~Dhariwal.
\newblock Glow: Generative flow with invertible 1x1 convolutions.
\newblock {\em Advances in Neural Information Processing Systems}, 31, 2018.

\bibitem{krizhevsky2009cifar-10}
A.~Krizhevsky.
\newblock Learning multiple layers of features from tiny images.
\newblock {\em University of Toronto}, 05 2012.

\bibitem{lecun1998mnist}
Y.~LeCun, L.~Bottou, Y.~Bengio, and P.~Haffner.
\newblock Gradient-based learning applied to document recognition.
\newblock {\em Proceedings of the IEEE}, 86(11):2278--2324, 1998. \href{https://doi.org/10.1109/5.726791}
{doi: {{%
10\hspace{.1pt}\discretionary{.}{%
}{.}\hspace{.4pt}1109\discretionary{/}{%
}{/}5\hspace{.1pt}\discretionary{.}{%
}{.}\hspace{.4pt}726791}}}


\bibitem{li2021visual}
H.~Li, M.~Xu, Y.~Wang, H.~Wei, and H.~Qu.
\newblock A visual analytics approach to facilitate the proctoring of online exams.
\newblock In {\em Proceedings of the CHI Conference on Human Factors in Computing Systems}, pp. 1--17, 2021. \href{https://doi.org/10.1145/3411764.3445294}
{doi: {{%
10\hspace{.1pt}\discretionary{.}{%
}{.}\hspace{.4pt}1145\discretionary{/}{%
}{/}3411764\hspace{.1pt}\discretionary{.}{%
}{.}\hspace{.4pt}3445294}}}


\bibitem{liu2019bZIP73}
C.~Liu, M.~R. Schl\"{a}ppi, B.~Mao, W.~Wang, A.~Wang, and C.~Chu.
\newblock The bzip73 transcription factor controls rice cold tolerance at the reproductive stage.
\newblock {\em Plant Biotechnology Journal}, 17(9):1834–1849, 2019. \href{https://doi.org/10.1111/pbi.13104}
{doi: {{%
10\hspace{.1pt}\discretionary{.}{%
}{.}\hspace{.4pt}1111\discretionary{/}{%
}{/}pbi\hspace{.1pt}\discretionary{.}{%
}{.}\hspace{.4pt}13104}}}


\bibitem{liu2019interative}
S.~Liu, C.~Chen, Y.~Lu, F.~Ouyang, and B.~Wang.
\newblock An interactive method to improve crowdsourced annotations.
\newblock {\em IEEE Transactions on Visualization and Computer Graphics}, 25(1):235--245, 2019. \href{https://doi.org/10.1109/TVCG.2018.2864843}
{doi: {{%
10\hspace{.1pt}\discretionary{.}{%
}{.}\hspace{.4pt}1109\discretionary{/}{%
}{/}TVCG\hspace{.1pt}\discretionary{.}{%
}{.}\hspace{.4pt}2018\hspace{.1pt}\discretionary{.}{%
}{.}\hspace{.4pt}2864843}}}


\bibitem{Liu_Liang_Gitter_2019}
S.~Liu, Y.~Liang, and A.~Gitter.
\newblock Loss-balanced task weighting to reduce negative transfer in multi-task learning.
\newblock In {\em Proceedings of the AAAI Conference on Artificial Intelligence}, pp. 9977--9978, 2019. \href{https://doi.org/10.1609/aaai.v33i01.33019977}
{doi: {{%
10\hspace{.1pt}\discretionary{.}{%
}{.}\hspace{.4pt}1609\discretionary{/}{%
}{/}aaai\hspace{.1pt}\discretionary{.}{%
}{.}\hspace{.4pt}v33i01\hspace{.1pt}\discretionary{.}{%
}{.}\hspace{.4pt}33019977}}}


\bibitem{lv2020resequencing}
Q.~Lv, W.~Li, Z.~Sun, N.~Ouyang, X.~Jing, Q.~He, J.~Wu, J.~Zheng, J.~Zheng, S.~Tang, et~al.
\newblock Resequencing of 1,143 indica rice accessions reveals important genetic variations and different heterosis patterns.
\newblock {\em Nature Communications}, 11(1):4778, 2020. \href{https://doi.org/10.1038/s41467-020-18608-0}
{doi: {{%
10\hspace{.1pt}\discretionary{.}{%
}{.}\hspace{.4pt}1038\discretionary{/}{%
}{/}s41467\discretionary{%
}{-}{-}020\discretionary{%
}{-}{-}18608\discretionary{%
}{-}{-}0}}}


\bibitem{l2022multi}
S.~L'Yi and N.~Gehlenborg.
\newblock Multi-view design patterns and responsive visualization for genomics data.
\newblock {\em IEEE Transactions on Visualization and Computer Graphics}, 29(1):559--569, 2022. \href{https://doi.org/10.1109/TVCG.2022.3209398}
{doi: {{%
10\hspace{.1pt}\discretionary{.}{%
}{.}\hspace{.4pt}1109\discretionary{/}{%
}{/}TVCG\hspace{.1pt}\discretionary{.}{%
}{.}\hspace{.4pt}2022\hspace{.1pt}\discretionary{.}{%
}{.}\hspace{.4pt}3209398}}}


\bibitem{lyi2021gosling}
S.~LYi, Q.~Wang, F.~Lekschas, and N.~Gehlenborg.
\newblock Gosling: A grammar-based toolkit for scalable and interactive genomics data visualization.
\newblock {\em IEEE Transactions on Visualization and Computer Graphics}, 28(1):140--150, 2021. \href{https://doi.org/10.1109/TVCG.2021.3114876}
{doi: {{%
10\hspace{.1pt}\discretionary{.}{%
}{.}\hspace{.4pt}1109\discretionary{/}{%
}{/}TVCG\hspace{.1pt}\discretionary{.}{%
}{.}\hspace{.4pt}2021\hspace{.1pt}\discretionary{.}{%
}{.}\hspace{.4pt}3114876}}}


\bibitem{lyu2024supercomputer}
F.~Lyu, C.~Chen, J.~Zhang, X.~Feng, and Z.~Tang.
\newblock Visualization for supercomputer system: A survey.
\newblock {\em Journal of Computer-Aided Design and Computer Graphics}, 36(3):321--335, 2024. \href{https://doi.org/10.3724/SP.J.1089.2024.2023-00791}
{doi: {{%
10\hspace{.1pt}\discretionary{.}{%
}{.}\hspace{.4pt}3724\discretionary{/}{%
}{/}SP\hspace{.1pt}\discretionary{.}{%
}{.}\hspace{.4pt}J\hspace{.1pt}\discretionary{.}{%
}{.}\hspace{.4pt}1089\hspace{.1pt}\discretionary{.}{%
}{.}\hspace{.4pt}2024\hspace{.1pt}\discretionary{.}{%
}{.}\hspace{.4pt}2023\discretionary{%
}{-}{-}00791}}}


\bibitem{mishra2025promptaid}
A.~Mishra, B.~Danzy, U.~Soni, A.~Arunkumar, J.~Huang, B.~C. Kwon, and C.~Bryan.
\newblock {PromptAid}: Visual prompt exploration, perturbation, testing and iteration for large language models.
\newblock {\em IEEE Transactions on Visualization and Computer Graphics}, 2025.
\newblock to be published. \href{https://doi.org/10.1109/tvcg.2025.3535332}
{doi: {{%
10\hspace{.1pt}\discretionary{.}{%
}{.}\hspace{.4pt}1109\discretionary{/}{%
}{/}tvcg\hspace{.1pt}\discretionary{.}{%
}{.}\hspace{.4pt}2025\hspace{.1pt}\discretionary{.}{%
}{.}\hspace{.4pt}3535332}}}


\bibitem{muller2021integrated}
J.~Muller, L.~Garrison, P.~Ulbrich, S.~Schreiber, S.~Bruckner, H.~Hauser, and S.~Oeltze-Jafra.
\newblock Integrated dual analysis of quantitative and qualitative high-dimensional data.
\newblock {\em IEEE Transactions on Visualization and Computer Graphics}, 27(6):2953--2966, 2021. \href{https://doi.org/10.1109/TVCG.2021.3056424}
{doi: {{%
10\hspace{.1pt}\discretionary{.}{%
}{.}\hspace{.4pt}1109\discretionary{/}{%
}{/}TVCG\hspace{.1pt}\discretionary{.}{%
}{.}\hspace{.4pt}2021\hspace{.1pt}\discretionary{.}{%
}{.}\hspace{.4pt}3056424}}}


\bibitem{nguyen2016visual}
Q.~V. Nguyen, N.~H. Khalifa, P.~Alzamora, A.~Gleeson, D.~Catchpoole, P.~J. Kennedy, and S.~Simoff.
\newblock Visual analytics of complex genomics data to guide effective treatment decisions.
\newblock {\em Journal of Imaging}, 2(4):29, 2016. \href{https://doi.org/10.3390/jimaging2040029}
{doi: {{%
10\hspace{.1pt}\discretionary{.}{%
}{.}\hspace{.4pt}3390\discretionary{/}{%
}{/}jimaging2040029}}}


\bibitem{nusrat2019tasks}
S.~Nusrat, T.~Harbig, and N.~Gehlenborg.
\newblock Tasks, techniques, and tools for genomic data visualization.
\newblock {\em Computer Graphics Forum}, 38(3):781--805, 2019. \href{https://doi.org/doi.org/10.1111/cgf.13727}
{doi: {{%
doi\hspace{.1pt}\discretionary{.}{%
}{.}\hspace{.4pt}org\discretionary{/}{%
}{/}10\hspace{.1pt}\discretionary{.}{%
}{.}\hspace{.4pt}1111\discretionary{/}{%
}{/}cgf\hspace{.1pt}\discretionary{.}{%
}{.}\hspace{.4pt}13727}}}


\bibitem{pandey2022genorec}
A.~Pandey, S.~L'Yi, Q.~Wang, M.~A. Borkin, and N.~Gehlenborg.
\newblock {GenoREC}: a recommendation system for interactive genomics data visualization.
\newblock {\em IEEE Transactions on Visualization and Computer Graphics}, 29(1):570--580, 2022. \href{https://doi.org/10.1109/TVCG.2022.3209407}
{doi: {{%
10\hspace{.1pt}\discretionary{.}{%
}{.}\hspace{.4pt}1109\discretionary{/}{%
}{/}TVCG\hspace{.1pt}\discretionary{.}{%
}{.}\hspace{.4pt}2022\hspace{.1pt}\discretionary{.}{%
}{.}\hspace{.4pt}3209407}}}


\bibitem{radford2021clip}
A.~Radford, J.~W. Kim, C.~Hallacy, A.~Ramesh, G.~Goh, S.~Agarwal, G.~Sastry, A.~Askell, P.~Mishkin, J.~Clark, G.~Krueger, and I.~Sutskever.
\newblock Learning transferable visual models from natural language supervision.
\newblock In {\em Proceedings of the International Conference on Machine Learning}, vol. 139, pp. 8748--8763, 2021.

\bibitem{rauber2018projections}
P.~E. Rauber, A.~X. Falcao, and A.~C. Telea.
\newblock Projections as visual aids for classification system design.
\newblock {\em Information Visualization}, 17(4):282--305, 2018. \href{https://doi.org/10.1177/1473871617713337}
{doi: {{%
10\hspace{.1pt}\discretionary{.}{%
}{.}\hspace{.4pt}1177\discretionary{/}{%
}{/}1473871617713337}}}


\bibitem{rombach2022high}
R.~Rombach, A.~Blattmann, D.~Lorenz, P.~Esser, and B.~Ommer.
\newblock High-resolution image synthesis with latent diffusion models.
\newblock In {\em Proceedings of the IEEE/CVF Conference on Computer Vision and Pattern Recognition}, pp. 10684--10695, 2022. \href{https://doi.org/10.1109/CVPR52688.2022.01042}
{doi: {{%
10\hspace{.1pt}\discretionary{.}{%
}{.}\hspace{.4pt}1109\discretionary{/}{%
}{/}CVPR52688\hspace{.1pt}\discretionary{.}{%
}{.}\hspace{.4pt}2022\hspace{.1pt}\discretionary{.}{%
}{.}\hspace{.4pt}01042}}}


\bibitem{self2016bridging}
J.~Z. Self, R.~K. Vinayagam, J.~T. Fry, and C.~North.
\newblock Bridging the gap between user intention and model parameters for human-in-the-loop data analytics.
\newblock In {\em Proceedings of the Workshop on Human-In-the-Loop Data Analytics}, pp. 1--6, 2016. \href{https://doi.org/10.1145/2939502.2939505}
{doi: {{%
10\hspace{.1pt}\discretionary{.}{%
}{.}\hspace{.4pt}1145\discretionary{/}{%
}{/}2939502\hspace{.1pt}\discretionary{.}{%
}{.}\hspace{.4pt}2939505}}}


\bibitem{tang2024jcvi}
H.~Tang, V.~Krishnakumar, X.~Zeng, Z.~Xu, A.~Taranto, J.~S. Lomas, Y.~Zhang, Y.~Huang, Y.~Wang, W.~C. Yim, et~al.
\newblock {JCVI}: A versatile toolkit for comparative genomics analysis.
\newblock {\em Imeta}, 3(4):e211, 2024. \href{https://doi.org/10.1002/imt2.211}
{doi: {{%
10\hspace{.1pt}\discretionary{.}{%
}{.}\hspace{.4pt}1002\discretionary{/}{%
}{/}imt2\hspace{.1pt}\discretionary{.}{%
}{.}\hspace{.4pt}211}}}


\bibitem{thorvaldsdottir2013igv}
H.~Thorvaldsd{\'o}ttir, J.~T. Robinson, and J.~P. Mesirov.
\newblock Integrative genomics viewer (igv): high-performance genomics data visualization and exploration.
\newblock {\em Briefings in Bioinformatics}, 14(2):178--192, 2013. \href{https://doi.org/10.1093/bib/bbs017}
{doi: {{%
10\hspace{.1pt}\discretionary{.}{%
}{.}\hspace{.4pt}1093\discretionary{/}{%
}{/}bib\discretionary{/}{%
}{/}bbs017}}}


\bibitem{turkay2016designing}
C.~Turkay, E.~Kaya, S.~Balcisoy, and H.~Hauser.
\newblock Designing progressive and interactive analytics processes for high-dimensional data analysis.
\newblock {\em IEEE Transactions on Visualization and Computer Graphics}, 23(1):131--140, 2016. \href{https://doi.org/10.1109/TVCG.2016.2598470}
{doi: {{%
10\hspace{.1pt}\discretionary{.}{%
}{.}\hspace{.4pt}1109\discretionary{/}{%
}{/}TVCG\hspace{.1pt}\discretionary{.}{%
}{.}\hspace{.4pt}2016\hspace{.1pt}\discretionary{.}{%
}{.}\hspace{.4pt}2598470}}}


\bibitem{turkay2014characterizing}
C.~Turkay, A.~Lex, M.~Streit, H.~Pfister, and H.~Hauser.
\newblock Characterizing cancer subtypes using dual analysis in caleydo stratomex.
\newblock {\em IEEE Computer Graphics and Applications}, 34(2):38--47, 2014. \href{https://doi.org/10.1109/MCG.2014.1}
{doi: {{%
10\hspace{.1pt}\discretionary{.}{%
}{.}\hspace{.4pt}1109\discretionary{/}{%
}{/}MCG\hspace{.1pt}\discretionary{.}{%
}{.}\hspace{.4pt}2014\hspace{.1pt}\discretionary{.}{%
}{.}\hspace{.4pt}1}}}


\bibitem{turkay2012representative}
C.~Turkay, A.~Lundervold, A.~J. Lundervold, and H.~Hauser.
\newblock Representative factor generation for the interactive visual analysis of high-dimensional data.
\newblock {\em IEEE Transactions on Visualization and Computer Graphics}, 18(12):2621--2630, 2012. \href{https://doi.org/10.1109/TVCG.2012.256}
{doi: {{%
10\hspace{.1pt}\discretionary{.}{%
}{.}\hspace{.4pt}1109\discretionary{/}{%
}{/}TVCG\hspace{.1pt}\discretionary{.}{%
}{.}\hspace{.4pt}2012\hspace{.1pt}\discretionary{.}{%
}{.}\hspace{.4pt}256}}}


\bibitem{turkay2014attribute}
C.~Turkay, A.~Slingsby, H.~Hauser, J.~Wood, and J.~Dykes.
\newblock Attribute signatures: Dynamic visual summaries for analyzing multivariate geographical data.
\newblock {\em IEEE Transactions on Visualization and Computer Graphics}, 20(12):2033--2042, 2014. \href{https://doi.org/10.1109/TVCG.2014.2346265}
{doi: {{%
10\hspace{.1pt}\discretionary{.}{%
}{.}\hspace{.4pt}1109\discretionary{/}{%
}{/}TVCG\hspace{.1pt}\discretionary{.}{%
}{.}\hspace{.4pt}2014\hspace{.1pt}\discretionary{.}{%
}{.}\hspace{.4pt}2346265}}}


\bibitem{astrid2023panva}
A.~van~den Brandt, E.~M. Jonkheer, D.-J.~M. van Workum, H.~van~de Wetering, S.~Smit, and A.~Vilanova.
\newblock {PanVA}: Pangenomic variant analysis.
\newblock {\em IEEE Transactions on Visualization and Computer Graphics}, 30(8):4895--4909, 2024. \href{https://doi.org/10.1109/TVCG.2023.3282364}
{doi: {{%
10\hspace{.1pt}\discretionary{.}{%
}{.}\hspace{.4pt}1109\discretionary{/}{%
}{/}TVCG\hspace{.1pt}\discretionary{.}{%
}{.}\hspace{.4pt}2023\hspace{.1pt}\discretionary{.}{%
}{.}\hspace{.4pt}3282364}}}


\bibitem{van2016exploring}
P.~van~der Corput and J.~J. van Wijk.
\newblock Exploring items and features with if, fi-tables.
\newblock {\em Computer Graphics Forum}, 35(3):31--40, 2016. \href{https://doi.org/10.1111/cgf.12879}
{doi: {{%
10\hspace{.1pt}\discretionary{.}{%
}{.}\hspace{.4pt}1111\discretionary{/}{%
}{/}cgf\hspace{.1pt}\discretionary{.}{%
}{.}\hspace{.4pt}12879}}}


\bibitem{van2008tsne}
L.~Van~der Maaten and G.~Hinton.
\newblock Visualizing data using t-sne.
\newblock {\em Journal of Machine Learning Research}, 9(11):2579--2605, 2008.

\bibitem{wang2023dnngp}
K.~Wang, M.~A. Abid, A.~Rasheed, J.~Crossa, S.~Hearne, and H.~Li.
\newblock {DNNGP}, a deep neural network-based method for genomic prediction using multi-omics data in plants.
\newblock {\em Molecular Plant}, 16(1):279--293, 2023. \href{https://doi.org/10.1016/j.molp.2022.11.004}
{doi: {{%
10\hspace{.1pt}\discretionary{.}{%
}{.}\hspace{.4pt}1016\discretionary{/}{%
}{/}j\hspace{.1pt}\discretionary{.}{%
}{.}\hspace{.4pt}molp\hspace{.1pt}\discretionary{.}{%
}{.}\hspace{.4pt}2022\hspace{.1pt}\discretionary{.}{%
}{.}\hspace{.4pt}11\hspace{.1pt}\discretionary{.}{%
}{.}\hspace{.4pt}004}}}


\bibitem{wang2023enabling}
Q.~Wang, X.~Liu, M.~Q. Liang, S.~L’Yi, and N.~Gehlenborg.
\newblock Enabling multimodal user interactions for genomics visualization creation.
\newblock In {\em Proceedings of the IEEE Visualization and Visual Analytics}, pp. 111--115, 2023. \href{https://doi.org/10.1109/VIS54172.2023.00031}
{doi: {{%
10\hspace{.1pt}\discretionary{.}{%
}{.}\hspace{.4pt}1109\discretionary{/}{%
}{/}VIS54172\hspace{.1pt}\discretionary{.}{%
}{.}\hspace{.4pt}2023\hspace{.1pt}\discretionary{.}{%
}{.}\hspace{.4pt}00031}}}


\bibitem{wang2023hetvis}
X.~Wang, W.~Chen, J.~Xia, Z.~Wen, R.~Zhu, and T.~Schreck.
\newblock {HetVis}: A visual analysis approach for identifying data heterogeneity in horizontal federated learning.
\newblock {\em IEEE Transactions on Visualization and Computer Graphics}, 29(1):310--319, 2023. \href{https://doi.org/10.1109/TVCG.2022.3209347}
{doi: {{%
10\hspace{.1pt}\discretionary{.}{%
}{.}\hspace{.4pt}1109\discretionary{/}{%
}{/}TVCG\hspace{.1pt}\discretionary{.}{%
}{.}\hspace{.4pt}2022\hspace{.1pt}\discretionary{.}{%
}{.}\hspace{.4pt}3209347}}}


\bibitem{wang2016topicpanorama}
X.~Wang, S.~Liu, J.~Liu, J.~Chen, J.~Zhu, and B.~Guo.
\newblock {TopicPanorama}: A full picture of relevant topics.
\newblock {\em IEEE Transactions on Visualization and Computer Graphics}, 22(12):2508–2521, 2016. \href{https://doi.org/10.1109/tvcg.2016.2515592}
{doi: {{%
10\hspace{.1pt}\discretionary{.}{%
}{.}\hspace{.4pt}1109\discretionary{/}{%
}{/}tvcg\hspace{.1pt}\discretionary{.}{%
}{.}\hspace{.4pt}2016\hspace{.1pt}\discretionary{.}{%
}{.}\hspace{.4pt}2515592}}}


\bibitem{wei2021quantitative}
X.~Wei, J.~Qiu, K.~Yong, J.~Fan, Q.~Zhang, H.~Hua, J.~Liu, Q.~Wang, K.~M. Olsen, B.~Han, et~al.
\newblock A quantitative genomics map of rice provides genetic insights and guides breeding.
\newblock {\em Nature Genetics}, 53(2):243--253, 2021. \href{https://doi.org/10.1038/s41588-020-00769-9}
{doi: {{%
10\hspace{.1pt}\discretionary{.}{%
}{.}\hspace{.4pt}1038\discretionary{/}{%
}{/}s41588\discretionary{%
}{-}{-}020\discretionary{%
}{-}{-}00769\discretionary{%
}{-}{-}9}}}


\bibitem{wold1987pca}
S.~Wold, K.~Esbensen, and P.~Geladi.
\newblock Principal component analysis.
\newblock {\em Chemometrics and Intelligent Laboratory Systems}, 2(1-3):37--52, 1987. \href{https://doi.org/10.1007/springerreference_84147}
{doi: {{%
10\hspace{.1pt}\discretionary{.}{%
}{.}\hspace{.4pt}1007\discretionary{/}{%
}{/}springerreference\_84147}}}


\bibitem{xia2022cdr}
J.~Xia, L.~Huang, W.~Lin, X.~Zhao, J.~Wu, Y.~Chen, Y.~Zhao, and W.~Chen.
\newblock Interactive visual cluster analysis by contrastive dimensionality reduction.
\newblock {\em IEEE Transactions on Visualization and Computer Graphics}, 29(1):734--744, 2022. \href{https://doi.org/10.1109/tvcg.2022.3209423}
{doi: {{%
10\hspace{.1pt}\discretionary{.}{%
}{.}\hspace{.4pt}1109\discretionary{/}{%
}{/}tvcg\hspace{.1pt}\discretionary{.}{%
}{.}\hspace{.4pt}2022\hspace{.1pt}\discretionary{.}{%
}{.}\hspace{.4pt}3209423}}}


\bibitem{xia2024parallel}
J.~Xia, L.~Huang, Y.~Sun, Z.~Deng, X.~L. Zhang, and M.~Zhu.
\newblock A parallel framework for streaming dimensionality reduction.
\newblock {\em IEEE Transactions on Visualization and Computer Graphics}, 30(1):142--152, 2024. \href{https://doi.org/10.1109/TVCG.2023.3326515}
{doi: {{%
10\hspace{.1pt}\discretionary{.}{%
}{.}\hspace{.4pt}1109\discretionary{/}{%
}{/}TVCG\hspace{.1pt}\discretionary{.}{%
}{.}\hspace{.4pt}2023\hspace{.1pt}\discretionary{.}{%
}{.}\hspace{.4pt}3326515}}}


\bibitem{xu2021genomic}
Y.~Xu, K.~Ma, Y.~Zhao, X.~Wang, K.~Zhou, G.~Yu, C.~Li, P.~Li, Z.~Yang, C.~Xu, et~al.
\newblock Genomic selection: A breakthrough technology in rice breeding.
\newblock {\em The Crop Journal}, 9(3):669--677, 2021. \href{https://doi.org/10.1016/j.cj.2021.03.008}
{doi: {{%
10\hspace{.1pt}\discretionary{.}{%
}{.}\hspace{.4pt}1016\discretionary{/}{%
}{/}j\hspace{.1pt}\discretionary{.}{%
}{.}\hspace{.4pt}cj\hspace{.1pt}\discretionary{.}{%
}{.}\hspace{.4pt}2021\hspace{.1pt}\discretionary{.}{%
}{.}\hspace{.4pt}03\hspace{.1pt}\discretionary{.}{%
}{.}\hspace{.4pt}008}}}


\bibitem{yang2024foundation}
W.~Yang, M.~Liu, Z.~Wang, and S.~Liu.
\newblock Foundation models meet visualizations: Challenges and opportunities.
\newblock {\em Computational Visual Media}, 10(3):399--424, 2024.

\bibitem{ye2025modalchorus}
Y.~Ye, S.~Xiao, X.~Zeng, and W.~Zeng.
\newblock {ModalChorus}: Visual probing and alignment of multi-modal embeddings via modal fusion map.
\newblock {\em IEEE Transactions on Visualization and Computer Graphics}, 31(1):294–304, 2025. \href{https://doi.org/10.1109/tvcg.2024.3456387}
{doi: {{%
10\hspace{.1pt}\discretionary{.}{%
}{.}\hspace{.4pt}1109\discretionary{/}{%
}{/}tvcg\hspace{.1pt}\discretionary{.}{%
}{.}\hspace{.4pt}2024\hspace{.1pt}\discretionary{.}{%
}{.}\hspace{.4pt}3456387}}}


\bibitem{yuan2017progress}
L.~Yuan.
\newblock Progress in super-hybrid rice breeding.
\newblock {\em The Crop Journal}, 5(2):100--102, 2017. \href{https://doi.org/10.1016/j.cj.2017.02.001}
{doi: {{%
10\hspace{.1pt}\discretionary{.}{%
}{.}\hspace{.4pt}1016\discretionary{/}{%
}{/}j\hspace{.1pt}\discretionary{.}{%
}{.}\hspace{.4pt}cj\hspace{.1pt}\discretionary{.}{%
}{.}\hspace{.4pt}2017\hspace{.1pt}\discretionary{.}{%
}{.}\hspace{.4pt}02\hspace{.1pt}\discretionary{.}{%
}{.}\hspace{.4pt}001}}}


\bibitem{yuan2013dimension}
X.~Yuan, D.~Ren, Z.~Wang, and C.~Guo.
\newblock Dimension projection matrix/tree: Interactive subspace visual exploration and analysis of high dimensional data.
\newblock {\em IEEE Transactions on Visualization and Computer Graphics}, 19(12):2625--2633, 2013. \href{https://doi.org/10.1109/TVCG.2013.150}
{doi: {{%
10\hspace{.1pt}\discretionary{.}{%
}{.}\hspace{.4pt}1109\discretionary{/}{%
}{/}TVCG\hspace{.1pt}\discretionary{.}{%
}{.}\hspace{.4pt}2013\hspace{.1pt}\discretionary{.}{%
}{.}\hspace{.4pt}150}}}


\bibitem{zhao2019featureexplorer}
J.~Zhao, M.~Karimzadeh, A.~Masjedi, T.~Wang, X.~Zhang, M.~M. Crawford, and D.~S. Ebert.
\newblock {FeatureExplorer}: Interactive feature selection and exploration of regression models for hyperspectral images.
\newblock In {\em Proceedings of the IEEE Visualization Conference}, pp. 161--165, 2019. \href{https://doi.org/10.1109/VISUAL.2019.8933619}
{doi: {{%
10\hspace{.1pt}\discretionary{.}{%
}{.}\hspace{.4pt}1109\discretionary{/}{%
}{/}VISUAL\hspace{.1pt}\discretionary{.}{%
}{.}\hspace{.4pt}2019\hspace{.1pt}\discretionary{.}{%
}{.}\hspace{.4pt}8933619}}}


\bibitem{zhou2025hierarchical}
Y.~Zhou, C.~Chen, Z.~Shen, J.~Zhu, J.~Chen, W.~Yang, and S.~Liu.
\newblock Hierarchical fuzzy-cluster-aware grid layout for large-scale data.
\newblock {\em IEEE Transactions on Visualization and Computer Graphics}, 2025.
\newblock to be published. \href{https://doi.org/10.1109/TVCG.2025.3566558}
{doi: {{%
10\hspace{.1pt}\discretionary{.}{%
}{.}\hspace{.4pt}1109\discretionary{/}{%
}{/}TVCG\hspace{.1pt}\discretionary{.}{%
}{.}\hspace{.4pt}2025\hspace{.1pt}\discretionary{.}{%
}{.}\hspace{.4pt}3566558}}}


\end{thebibliography}

\end{document}


\maketitle

\section*{Appendix A: Derivation of Eq. (5)}
In the dual analysis framework, taking modifying $S$ to $S'$ as an example, the updating of X is formulated as:
\begin{equation}
    \begin{aligned}
    \label{eq:x_update}
        X' & = &  \underset{X} {\arg\min} & & &  \|p_s(X) - S'\|^2 \\
           & & \mathrm{s.t.}  & & &  p_s = \underset{p_s}{\arg\min}\, O_s(p_s, X).
    \end{aligned}
    \tag{3}
\end{equation}
When Multidimensional Scaling (MDS) is used as the projection method, since MDS directly provides the projection results rather than relying on a projection function, Eq.~(\ref{eq:x_update}) is transformed into:
\[
    \begin{aligned}
    \label{eq:x_update_mds}
        X' &=& \underset{X}{\arg\min} & & & \sum_{i} {(s_i - s'_i)^2}  \\
           & &\mathrm{s.t.} & & & S = \underset{S}{\arg\min}\, \sum_{i,j} \left\| \sqrt{(x_i - x_j)^2} - \sqrt{(s_i - s_j)^2} \right\|^2.
    \end{aligned}
\]
Assuming that ${X}''=\{x_1'',x_2'',..\}$ represents the optimal solution for the optimization problem:
\[
\underset{X}{\arg\min}\, \sum_{i,j} \left\| \sqrt{(x_i - x_j)^2} - \sqrt{(s'_i - s'_j)^2} \right\|^2.
\]
Then, in Eq.~(\ref{eq:x_update}), let $x_i=x_i''$ and $s_i=s_i'$ for all $i$.
This choice of values guarantees that the constraint in Eq.~(\ref{eq:x_update}) is satisfied, while the objective in Eq.~(\ref{eq:x_update}) is minimized (\ie, zero).
It means that the optimization problem above is equal to optimizing Eq.~(\ref{eq:x_update}).
When \( X' = XW \), the optimization problem above is transformed into Eq.~(\ref{eq:sirius}):
\begin{equation}
\begin{aligned}
    \label{eq:sirius}
    W &= \underset{W}{\arg\min} \sum_{i,j} \left\| \sqrt{((XW)_i - (XW)_j)^2} - \sqrt{(s'_i - s'_j)^2} \right\|^2 \\
      &= \underset{W}{\arg\min} \sum_{i,j} \left\| \sqrt{(x_i W - x_j W)^2} - \sqrt{(s'_i - s'_j)^2} \right\|^2.
\end{aligned}
\tag{5}
\end{equation}

\clearpage
\section*{Appendix B: Proof of Theorem 1}
\newtheorem{theorem}{Theorem}
\begin{theorem}
\label{theorem:opt}
Using an invertible neural network as the projection function $p_s(\cdot)$. When $S$ is modified to $S'$, let $X'_{\text{opt}}$ be the optimal updated X obtained by optimizing Eq.~(\ref{eq:x_update}).
Then we have
$$X'_{\text{opt}} = p_s^{-1}(S').$$
\end{theorem}
\begin{proof}
Since $ p_s $ is invertible, there exists a unique $ X = p_s^{-1}(S') $ such that $ p_s(X) = S' $. Then we obtain
$$
\| p_s(p_s^{-1}(S')) - S' \|^2 = \| S' - S' \|^2 = 0.
$$
This is the global minimum of $ \| p_s(X) - S' \|^2 $, which is non-negative and zero only when $ p_s(X) = S' $. 
Therefore, $X'_{\text{opt}} = p_s^{-1}(S')$.
\end{proof}

\section*{Appendix C: Proof of Theorem 2}
\begin{theorem}
\label{theorem:comparison}
Using an invertible neural network as the projection method $p_s(\cdot)$.
When $S$ is modified to $S'$, let $X'_{\text{SIRIUS}}$ be the updated X obtained by SIRIUS, and $X'_{\text{inv}}$ be the updated X obtained by our method.
Then we have 
$$\|p_s(X'_{inv}) - S'\|^2 \leq \|p_s(X'_{SIRIUS}) - S'\|^2.$$
\end{theorem}
\begin{proof}
By Theorem~\ref{theorem:opt}, when using an invertible neural network $p_s(\cdot)$, the optimal updated $X'_{\text{inv}}$ satisfies $X'_{\text{inv}} = p_s^{-1}(S')$, and thus $p_s(X'_{\text{inv}}) = S'$, implying
$$\|p_s(X'_{\text{inv}}) - S'\|^2 = 0.$$
Since $\|p_s(X'_{\text{SIRIUS}}) - S'\|^2 \geq 0$ for any $X'_{\text{SIRIUS}}$ produced by SIRIUS, it follows that $\|p_s(X'_{\text{inv}}) - S'\|^2 \leq \|p_s(X'_{\text{SIRIUS}}) - S'\|^2$.
\end{proof}

\clearpage
\section*{Appendix D: Neural Network Structure of Our Method}
\subsection*{D.1 Network structure}
\begin{figure}
    \centering
    \includegraphics[width=\linewidth]{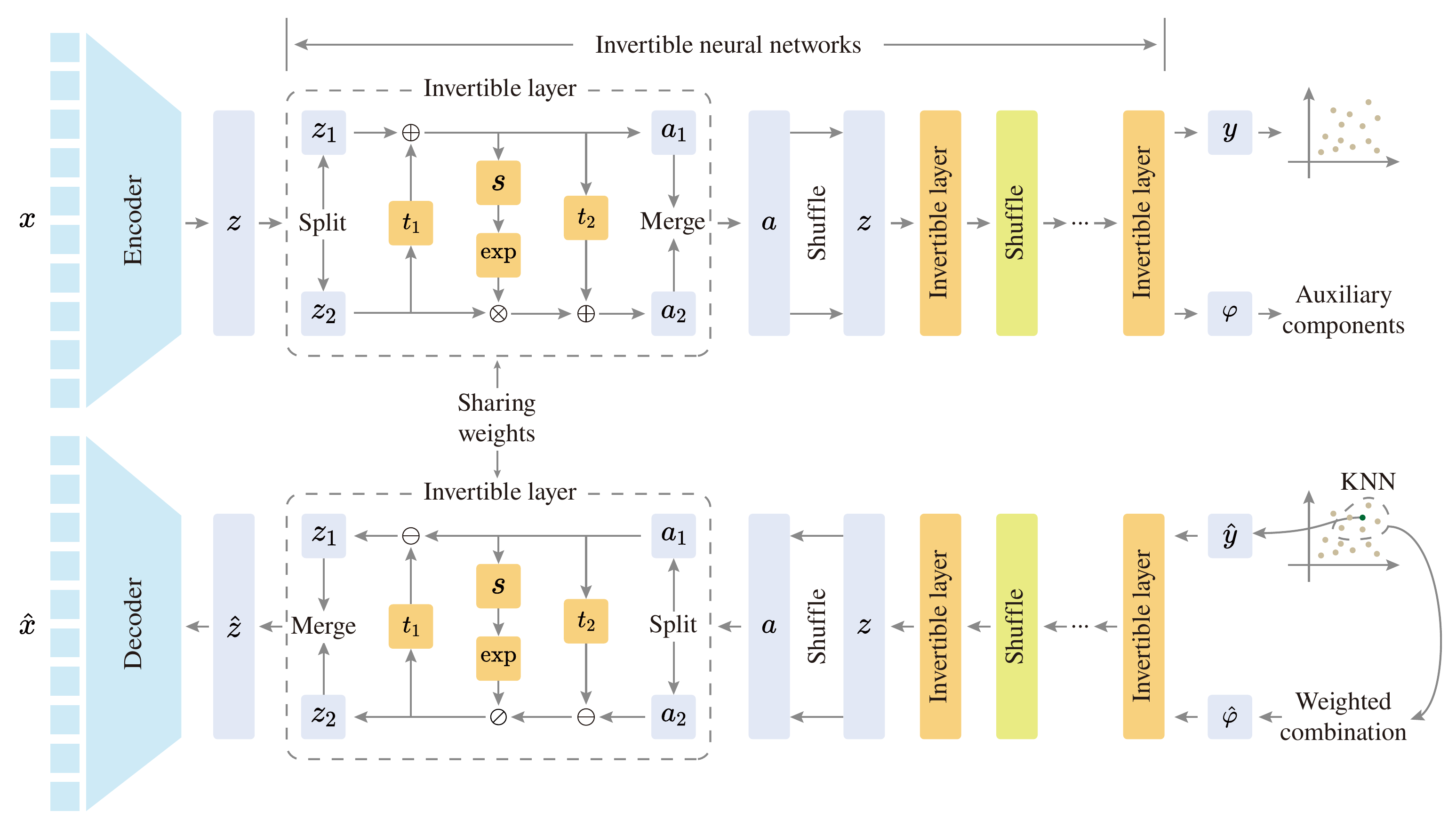}
    \caption{Neural network structure of our method.}
    \label{fig:model}
\end{figure}

\pcheng{Fig.~\ref{fig:model} illustrates our method’s neural network, integrating invertible neural networks (INNs) within an autoencoder. The autoencoder maps input $x$ to latent representation $z$, which INNs transform into a 2D projection $y$ and auxiliary component $\varphi$. Given a new 2D input $\hat{y}$, INNs compute the auxiliary component $\hat{\varphi}$ and invert them to $\hat{z}$, enabling reconstruction of $\hat{x}$.}

\myparagraph{\pcheng{Invertible layer.}}
\pcheng{INNs employ affine coupling layers (ACLs)~\cite{dinh2016density}, where each layer splits input $z$ into $(z_1, z_2)$ and outputs $(a_1, a_2)$. While standard ACLs set $a_1=z_1$, we enhance this through an additive transformation inspired by InvVis~\cite{ye2023invvis}. The forward transformation is expressed as:
\[
\begin{aligned}
    a_1 &= z_1 + t_1(z_2), \\
    a_2 &= z_2 \otimes \exp(s(a_1)) + t_2(a_1).
\end{aligned}
\]
The inverse is:
\[
\begin{aligned}
    z_2 &= (a_2 - t_2(a_1)) \otimes \exp(-s(a_1)), \\
    z_1 &= a_1 - t_1(z_2).
\end{aligned}
\]
Here, $\otimes$ is the Hadamard product, and $s(\cdot)$, $t_1(\cdot)$, $t_2(\cdot)$ are transformation blocks, implemented with fully connected layers.}

\pcheng{To avoid direct alignment between the first two dimensions of input $z$ and the output $y$, we apply shuffle layers inspired by RealNVP~\cite{dinh2016density} with a fixed random permutation $\pi$ between invertible layers.}

\myparagraph{\pcheng{Projection and reconstruction.}}
\pcheng{INNs continuously reduce the dimensionality of $z_1$ while increasing that of $z_2$. In the final layer, $a_1$ becomes the 2D projection $y$, and $a_2$ is $\varphi$. For a new $\hat{y}$, $\hat{\varphi}$ is computed using k-nearest neighbors (KNN) from training data:
\[
\hat{\varphi} = \frac{\sum_{i=1}^{k} \frac{1}{d(\hat{y}, y_{i})} \cdot \varphi_{i}}{\sum_{i=1}^{k} \frac{1}{d(\hat{y}, y_{i})}},
\]
where $d(\cdot)$ is Euclidean distance.}

\subsection*{D.2 Network configurations in the Dual Projection Experiment}

\myparagraph{\pcheng{Architecture details.}}
\pcheng{The autoencoder uses a four-layer fully connected network with batch normalization and ReLU activations for both encoder and decoder. The INNs comprise four invertible layers, with transformation blocks ($s(\cdot)$, $t_1(\cdot)$, $t_2(\cdot)$) as four-layer fully connected networks with batch normalization and ReLU. Hidden dimensions are 64 for MNIST and CIFAR-10, and 256 for the genomic dataset.}

\myparagraph{\pcheng{Training parameters.}}
\pcheng{We train with a batch size of $N/10$ (where $N$ is the dataset size), learning rate of 0.01, and 1000 epochs using the AdamW optimizer and cosine learning rate scheduler. For contrastive learning, we adopt the training protocol of CDR~\cite{xia2022cdr} but set the positive sample sampling range to $k=10$. The reconstruction loss weight $\lambda$ is set to 0.1.}

\clearpage
\section*{Appendix E: More Details of the Dual Projection Experiment}
\myparagraph{\pcheng{Measure definition.}} \pcheng{Trustworthiness ($T$) and Continuity ($C$) evaluate dimensionality reduction quality by assessing preservation of high-dimensional data structure in low-dimensional embeddings. $T$ ensures nearby points in high-dimensional space remain nearby in low-dimensional space. $C$ ensures points close in low-dimensional space are close in high-dimensional space.}

\pcheng{For $n$ data points and neighborhood size $k$:
\begin{equation}
T(k) = 1 - \frac{2}{nk (2n - 3k - 1)} \sum_{i=1}^n \sum_{j \in U_k(i)} (r(i,j) - k),
\end{equation}
where $U_k(i)$ is the set of points among $k$ nearest neighbors of point $i$ in low-dimensional space but not in high-dimensional space and $r(i,j)$ is the rank of point $j$ in point $i$'s high-dimensional neighbors.
}
\pcheng{
\begin{equation}
C(k) = 1 - \frac{2}{nk (2n - 3k - 1)} \sum_{i=1}^n \sum_{j \in V_k(i)} (\hat{r}(i,j) - k),
\end{equation}
where $V_k(i)$ is the set of points among $k$ nearest neighbors of point $i$ in high-dimensional space but not in low-dimensional space and $\hat{r}(i,j)$ is the rank of point $j$ in point $i$'s low-dimensional neighbors.
}

\myparagraph{\pcheng{The influence of k.}} 
\pcheng{In the dual projection experiment, we followed MFM~\cite{ye2025modalchorus} and set k as 30.
We also conducted an experiment to evaluate the influence of k. We tested k=20, 30, and 40 on the MNIST, CIFAR-10, and Genomic datasets.
The results are shown in Tab.~\ref{tab:trustworthiness_continuity}.
As we can see, our method consistently performs better than SIRIUS and is comparable to single projection methods.}

\begin{table}[ht]
\centering
\caption{\pcheng{Trustworthiness (T) and continuity (C) scores of single and dual projection methods for k=20, 30, and 40.}}
\label{tab:trustworthiness_continuity}
\pcheng{
\begin{subtable}{\textwidth}
\centering
\caption{\pcheng{Trustworthiness (T)}}
\begin{tabular}{c|c|ccc|ccc|ccc}
\hline
& \multirow{2}{*}{Methods} & \multicolumn{3}{c|}{MNIST} & \multicolumn{3}{c|}{CIFAR-10} & \multicolumn{3}{c}{Genomic dataset} \\
\cline{3-11}
& & T(20) & T(30) & T(40) & T(20) & T(30) & T(40) & T(20) & T(30) & T(40) \\
\hline
\multirow{2}{*}{Single} & t-SNE & \textbf{0.972} & \textbf{0.963} & \textbf{0.959} & \textbf{0.958} & \textbf{0.951} & \textbf{0.946} & \textbf{0.963} & \textbf{0.957} & \textbf{0.954} \\
& PCA & 0.735 & 0.735 & 0.735 & 0.784 & 0.784 & 0.785 & 0.855 & 0.856 & 0.858 \\
\hline
\multirow{2}{*}{Dual}& SIRIUS & 0.775 & 0.775 & 0.776 & 0.786 & 0.786 & 0.787 & 0.893 & 0.888 & 0.885 \\
& Ours   & \textbf{0.962} & \textbf{0.961} & \textbf{0.960} & \textbf{0.943} & \textbf{0.941} & \textbf{0.940} & \textbf{0.978} & \textbf{0.972} & \textbf{0.966} \\
\hline
\end{tabular}
\end{subtable}
}
\vspace{1em}
\pcheng{
\begin{subtable}{\textwidth}
\centering
\caption{\pcheng{Continuity (C)}}
\begin{tabular}{c|c|ccc|ccc|ccc}
\hline
& \multirow{2}{*}{Methods}  & \multicolumn{3}{c|}{MNIST} & \multicolumn{3}{c|}{CIFAR-10} & \multicolumn{3}{c}{Genomic dataset} \\
\cline{3-11}
& & C(20) & C(30) & C(40) & C(20) & C(30) & C(40) & C(20) & C(30) & C(40) \\
\hline
\multirow{2}{*}{Single} & t-SNE   & \textbf{0.958} & \textbf{0.950} & \textbf{0.945} & \textbf{0.962} & \textbf{0.958} & \textbf{0.955} & \textbf{0.956} & \textbf{0.948} & \textbf{0.943} \\
& PCA & 0.912 & 0.905 & 0.899 & 0.922 & 0.917 & 0.914 & 0.917 & 0.910 & 0.905 \\
\hline
\multirow{2}{*}{Dual}& SIRIUS & 0.892 & 0.888 & 0.886 & 0.897 & 0.895 & 0.893 & 0.922 & 0.919 & 0.918 \\
& Ours   & \textbf{0.951} & \textbf{0.944} & \textbf{0.938} & \textbf{0.953} & \textbf{0.949} & \textbf{0.946} & \textbf{0.959} & \textbf{0.950} & \textbf{0.944} \\
\hline
\end{tabular}
\end{subtable}
}
\end{table}

\clearpage
\section*{Appendix F: Full Qualitative Projection Results}
\begin{figure}
    \centering
    \includegraphics[width=\linewidth]{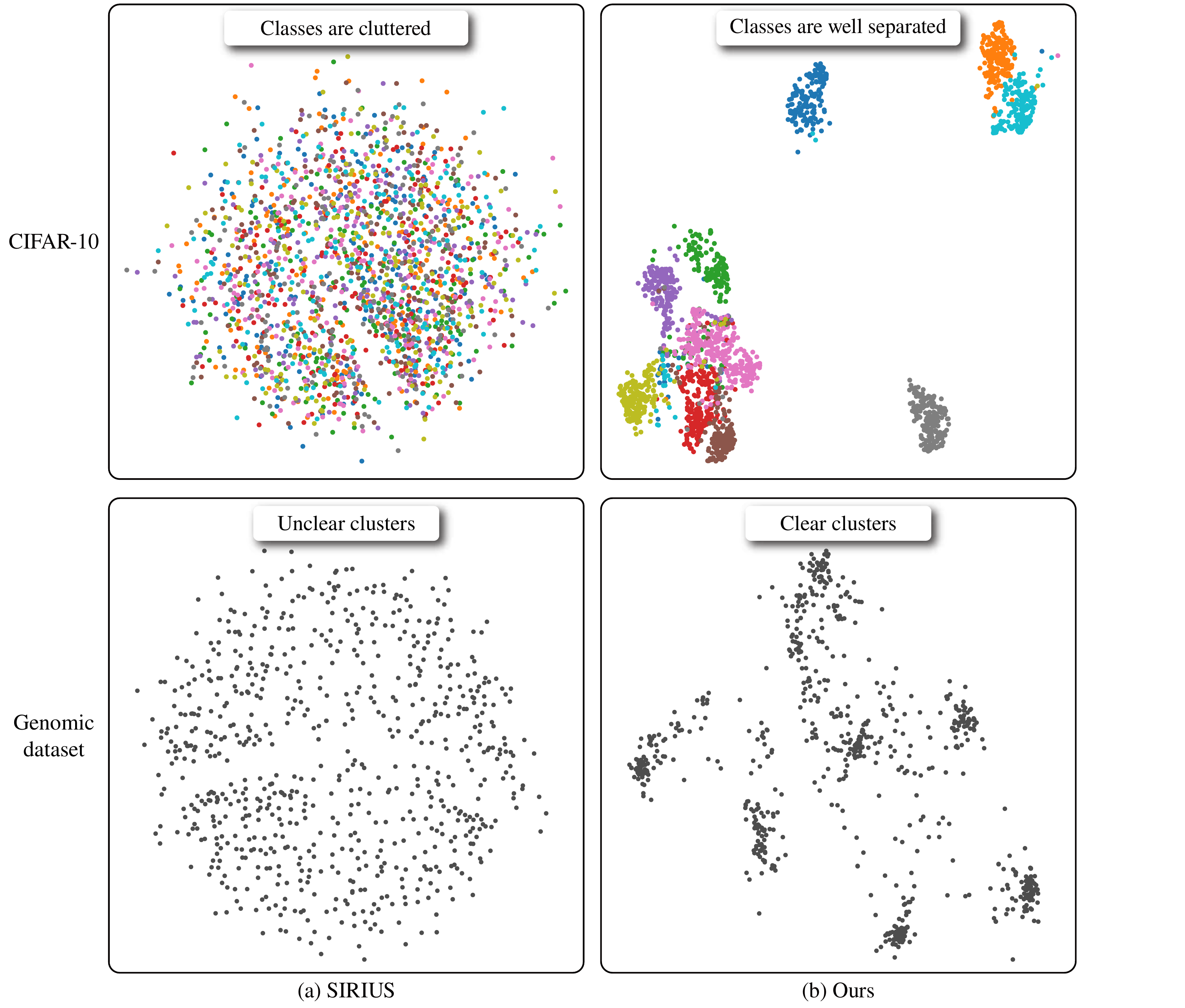}
    \caption{Visual comparison on the CIFAR-10 and genomic dataset.}
    \label{fig:cifar-gene}
\end{figure}

Fig.~\ref{fig:cifar-gene} presents the comparison between SIRIUS and our method on both the CIFAR-10~\cite{krizhevsky2009cifar-10} and genomic datasets, leading to similar conclusions as MNIST. On both datasets, SIRIUS produces densely clustered classes, hindering effective visual exploration. In contrast, our method consistently achieves clear class separation, highlighting its effectiveness in preserving neighborhood relationships.

\clearpage
\bibliographystyle{abbrv-doi-hyperref}

\bibliography{reference}